\documentclass[preprint,11pt]{elsarticle}
\usepackage{graphicx}
\usepackage{verbatim}
\usepackage{epsfig}
\usepackage{titlesec}
\usepackage{amsmath, amsfonts, amssymb}
\usepackage{mathrsfs}
\usepackage{xcolor}
\usepackage{rotating}
\usepackage{mathtools}
\usepackage{multirow}
\usepackage{amsthm}
\usepackage{esvect}
\usepackage{multicol}
\usepackage{subcaption}
\usepackage{algorithm}
\usepackage{algpseudocode}
\usepackage{geometry,longtable,booktabs, caption}
\usepackage{comment}
\usepackage{float}
\usepackage{enumerate}
\usepackage{bm}
\usepackage{natbib}
\usepackage{hyperref}
\usepackage{comment}
\setcitestyle{authoryear,open={(},close={)}} 

\titleformat{\subsection}
  {\normalfont\bfseries}
  {\thesubsection}
  {1em}
  {}
\titleformat{\subsubsection}
  {\normalfont\normalsize\bfseries}
  {\thesubsubsection}
  {1em}
  {}
\def\be{\begin{equation}}
\def\ee{\end{equation}}
\def\bq{\begin{eqnarray}}
\def\eq{\end{eqnarray}}

\makeatletter
\def\ps@pprintTitle{%
  \let\@oddhead\@empty
  \let\@evenhead\@empty
  \def\@oddfoot{\reset@font\hfil\thepage\hfil}
  \let\@evenfoot\@oddfoot
}
\makeatother

\begin{document}
\begin{frontmatter}
	
\title{A Temporal Multiplex Graph Neural Network for Systemic Risk Transmission in Global Banking}

\author{ Nneka Umeorah\footnotemark[1], Tolulope Fadina\footnotemark[2]}
\address{\footnotemark[1]Cardiff University; School of Mathematics; Cardiff CF24 4AG; United Kingdom\\
\footnotemark[2]Department of Mathematics, University of Illinois Urbana-Champaign, Champaign, USA [tfadina@illinois.edu]\\
\textbf{Correponding author}: Nneka Umeorah [umeorahn@cardiff.ac.uk] }

\begin{abstract}
This paper develops a unified framework for assessing systemic risk and identifying contagion channels in the global banking system using a Temporal Heterogeneous Multiplex Graph Neural Network. We construct a harmonised quarterly panel combining bank fundamentals, CDS spreads, and macroeconomic indicators, and represent these data as dynamic multiplex networks linking banks through financial similarity and liquidity co-movement, augmented with country-level macroeconomic relationships. The model integrates graph convolutional layers with recurrent GRU dynamics and incorporates a learnable fusion gate to capture time-varying reliance on alternative contagion channels. Empirical results show that the framework outperforms conventional econometric, machine learning, and graph-based benchmarks for short-term changes in CDS spreads. Beyond forecasting, we provide an interpretable framework to quantify bank-level systemic importance via stress testing, assess country-level spillovers under macroeconomic shocks, and uncover transmission pathways through edge perturbation analysis. Robustness tests confirm the stability of both predictive accuracy and systemic risk rankings.\\

\end{abstract}

\end{frontmatter}

\section{Introduction}
The transmission of credit risk across financial institutions is central to both asset pricing and financial stability. Events such as the Global Financial Crisis (GFC) and the COVID-19 shock have demonstrated that distress in one part of the financial system can propagate rapidly across institutions, amplifying systemic risk. Although existing research has made great progress in analysing these dynamics, fully characterising how spillovers arise, evolve, and interact across multiple channels remains a challenging problem for both policymakers and researchers.

To address these interconnected dynamics, network-based approaches have become increasingly important in systemic risk analysis; see \cite{bisias2012survey} for an overview of systemic risk analysis. Network models provide a natural framework for analysing financial contagion by explicitly representing interconnections between institutions. For instance, \cite{allen2000financial} shows how the structure of interbank linkages affects the propagation of shocks, \cite{acemoglu2015systemic} demonstrates that network topology plays a key role in determining whether shocks are absorbed or amplified, while \cite{fadinaKomla} finds that during the COVID-19 sell-off, US and EU collateral loan obligation markets showed limited contagion and interconnectedness with broader financial markets significantly affecting corporate bonds but showing little to no transmission to equities or government bonds. Further empirical implementations typically construct networks based on either observed bilateral exposures, such as interbank lending and payment obligations \citep{eisenberg2001systemic, upper2011simulation}, or statistical measures of dependence, including correlations and forecast-error variance decompositions \citep{billio2012econometric, diebold2014network}. These studies highlight the importance of interconnectedness in shaping systemic risk and provide a foundation for analysing financial contagion through network structures.

Network dependence is often modelled within linear econometric frameworks, such as panel regressions and vector autoregressions. These approaches are tractable and offer clear interpretability, making them widely used in empirical finance \citep{diebold2014network, fadinaKomla}. However, they typically impose fixed and homogeneous dependence structures and rely on linear dynamics, limiting their ability to capture nonlinear contagion effects and the time-varying nature of financial linkages. Moreover, connectedness measures derived from these models depend heavily on pre-specified system dynamics, which may not fully reflect the underlying transmission mechanisms observed during periods of financial stress. These limitations have motivated growing interest in more flexible, data-driven modelling frameworks.

Recent advances in machine learning aim to address these challenges by leveraging flexible frameworks that can model complex, nonlinear dependencies. In particular, graph-based approaches have gained traction by explicitly representing relational structure across observations; see \cite{Wusurvey21} and  \cite{WangZhang2022} for an overview. Multiplex (or multilayer) network representations provide a natural framework for modelling systems characterised by multiple, co-existing types of interactions \cite{de2013mathematical, xie2022systemic, gao2022systemic}. These approaches are especially relevant in financial systems, where interconnectedness evolves over time and transmission mechanisms may vary across market conditions. Recently, \cite{kumar2026regime} develops a regime-dependent GNN framework for volatility prediction, demonstrating that dynamically evolving graph structures can improve predictive performance across different market states. Similarly, \cite{guo2021tabgnn} shows that multiplex graph representations can improve learning performance by capturing multiple interaction channels simultaneously. Collectively, these studies highlight the potential of graph-based learning methods to model high-dimensional and nonlinear financial dependence structures beyond conventional econometric settings.

Despite these advances, some limitations remain. Many empirical implementations continue to rely on single-layer representations of interconnectedness, focusing either on exposures or on statistical dependence, and therefore fail to capture the inherently multidimensional nature of financial linkages, in which channels such as capital structure, liquidity conditions, and macroeconomic exposure interact but evolve differently over time. In addition, even when time variation is incorporated, dependence structures are often modelled in a restricted or smoothed manner, for example, through rolling-window estimation or slowly evolving estimation procedures. This may smooth abrupt changes in interconnectedness and could depend heavily on arbitrary window specifications \citep{diebold2012better, diebold2014network, liu2020analyzing}. Thus, this limits the ability to capture rapid, nonlinear shifts in interconnectedness, particularly during periods of financial stress.

To address these limitations, this paper proposes a Temporal Heterogeneous Multiplex Graph Neural Network (HMGNN) to model and forecast credit risk dynamics across financial institutions. The framework is heterogeneous because it jointly models multiple node types, namely banks and countries, along with multiple relation types that represent distinct channels of financial transmission. The framework also represents the financial system as a multi-layer network, where each layer captures a distinct channel of dependence: (i) a financial similarity layer based on banks' balance sheet characteristics, (ii) a time-varying liquidity co-movement layer capturing short-term funding dynamics, and (iii) a country–bank layer incorporating macroeconomic transmission. By integrating these layers into a unified temporal graph, the model captures both cross-sectional interdependencies and how they evolve over time.

To account for temporal dynamics, we combine the multiplex network with a recurrent architecture that learns how interbank dependencies evolve. A key feature of the model is a learnable fusion mechanism that dynamically weights the relative importance of different network layers, enabling the model to adapt across various economic regimes. This enables the framework to capture nonlinear, time-varying contagion channels that are difficult to model using traditional approaches. We apply the model to forecast changes in bank-level CDS spreads, a market-based measure of credit risk. Our empirical analysis focuses on large internationally active banks, including global Systemically Important Banks (SIBs), as these institutions occupy central positions in the global financial network. Their size, interconnectedness, and cross-border exposures make them key channels for propagating systemic risk and financial contagion.

Empirically, the proposed approach outperforms standard benchmarks, including panel regressions, vector autoregressions, and machine learning models, in terms of predictive accuracy. Beyond proposing a new GNN architecture, this paper develops a unified framework to simultaneously forecast bank credit risk, identify systemically important financial institutions, and decompose contagion into structural, liquidity, and macroeconomic transmission channels. By integrating predictive modelling with network-based interpretability, the proposed framework not only improves forecasting performance but also provides economically meaningful insights into how systemic risk propagates across the global banking system.

This paper makes three main contributions to the literature.
\begin{itemize}
    \item First, we develop a unified framework that simultaneously forecasts bank credit risk, identifies systemically important institutions, and analyses contagion within a single modelling framework, thereby linking prediction with systemic risk assessment.

\item Second, we propose a Temporal HMGNN that integrates structural financial similarity, dynamic liquidity co-movement, and country-level macroeconomic information via a learnable fusion mechanism, enabling the model to capture nonlinear, time-varying transmission dynamics.

\item Third, we provide an interpretable decomposition of systemic risk by quantifying the relative importance of structural, liquidity, and macroeconomic transmission channels, together with bank-level and country-level stress-testing analyses that reveal how shocks propagate through the global banking network.
\end{itemize}

The remainder of the paper is organised as follows. Section~\ref{sec2} reviews the related literature. Section~\ref{sec3} outlines the methodology, including the model architecture and problem formulation. Section~\ref{sec4} describes the data structure and empirical design and presents the benchmarking framework. Section~\ref{sec5} reports the empirical results, including statistical evaluation, systemic risk transmission analysis, interpretability, and robustness checks. Section~\ref{sec6} concludes the research.

\section{Related Works}\label{sec2}

\subsection{Network Models for Credit Default Swaps}
Credit default swaps provide a natural application of graph neural networks because the market consists of interconnected financial institutions whose relationships can be naturally represented as a network. This network structure has long been recognised in the CDS literature, even before graph learning methods became widely used.  \cite{chenwang2013} describes the CDS market itself as a network of banks, insurers, and hedge funds connected by exposure, with contagion-index measures capturing expected system-wide loss from a single default. One contagion-dynamics model also explicitly adds CDS contracts as a distinct, non-Markovian contagion channel layered on top of direct firm-bank exposures \cite{heise2012}.
\cite{giudici2018corisk} proposes a credit risk measurement model for sovereign CDS spreads that combines vector autoregressive regression with correlation networks, separating an idiosyncratic, country-specific component from a contemporaneous contagion component. \cite{chen2020network} combines conditional Granger causality and dynamic network analysis to examine interconnectedness and financial transmission channels within the European sovereign CDS market. Empirical results suggest short-term risk is volatile, while idiosyncratic connections behave as long-term pricing factors and transmit contagion \cite{chen2020network}.
\cite{Shen2020} constructs a sovereign CDS risk contagion network using generalised prediction error variance decomposition to understand spillover indices. The study identifies four risk contagion stages: neighbouring country contagion, Russia-Ukraine risk exacerbation, further outward spillover, and weakening but persistent spillover. \cite{Nguyen2024} combines CDS spreads with network and copula models to analyse the interdependencies between banks in the US and Europe. The models capture nonlinear tail dependence in CDS spreads, distinguishing between common macroeconomic shocks and contagion transmitted through financial networks. Empirical evidence suggests financial contagion operates through regional and global channels. At the regional level, both systematic and idiosyncratic contagion channels operate within Europe and the US individually; at the global level, systematic contagion dominates when considering both regions together. The dominance of systematic over idiosyncratic contagion in the EU and the US is consistent with the results in \cite{Ballester2016}. \cite{Nguyen2024} approach can be seen as an extension of \cite{OhPatton2018}, which used single-factor dynamic copulas for US firms, and \cite{Krup2020}, which applied time-varying factor copulas to US firms.

Despite this network-based framework, graph neural networks have only recently been applied directly to CDS. \cite{brogaardchen2024attention} implements an attention-based GNN where firms are nodes and idiosyncratic-volatility spillovers are directed edges, jointly capturing inter-firm contagion and firm features to predict CDS spreads. They remark that incorporating this network structure improves out-of-sample CDS spread prediction accuracy by more than 50\% over non-graph machine learning baselines that cannot use edge information. In a related work, \cite{Zandi2025} builds a dynamic, multilayer network over borrowers, each layer representing a distinct source of connection combined with a GNN-RNN architecture that lets default risk propagate across layers and evolve over time; empirical evidence suggests explicit multi-relational, temporal structure improves behavioural credit-scoring performance over borrower-level-only baselines. It is important to note that the \cite{brogaardchen2024attention} CDS model is single-relation and effectively static, while \cite{Zandi2025}'s multilayer model is not applied to CDS or firm-network contagion at all. 
\cite{Shu2026} applies GNNs to the interbank exposure networks that underlie CDS-driven systemic risk rather than to spread prediction itself. A related deep-learning contagion study builds time-varying hypergraphs over sector-level anomalous correlations, including sovereign CDS risk as one tracked participant, to capture multi-sector risk communities beyond pairwise correlation \cite{Akguller2026}, emphasising the CDS structure noted in \cite{heise2012}. GNN-based models have been shown to outperform traditional machine learning at classifying banks' systemic importance in simulated exposure networks, and other work integrates GNNs directly with interbank liability networks to compute Eisenberg--Noe-style systemic risk measures from graph-structured data \cite{Shu2026}. 

\subsection{Temporal Multiplex GNN}

Surveys on the application of GNN to finance, see for example, \cite{Wusurvey21} and \cite{WangZhang2022}, categorize graph construction techniques and summarize major GNN architectures into graph convolutional networks (GCNs) \cite{kipf2017semi}, graph attention networks \cite{brogaardchen2024attention}, GraphSAGE \cite{hamilton2017inductive}, heterogeneous GNNs \cite{Tan2022}, temporal GNNs \cite{Feng2019}, heterogeneous temporal GNN \cite{xiang2022temporal}, and hypergraph neural networks \cite{Sadek2025} and review their applications in finance. 

Recent developments have focused on jointly modelling temporal dependence and heterogeneous graph structures via heterogeneous temporal GNN (HTG). The HTG was proposed by \cite{fan2022heterogeneous} with successful applications in computational finance \cite{xiang2022temporal}, e-commerce \cite{liu2024kddc, xie2021learning, qiao2025gcal}, traffic network \cite{liu2024full, feng2022adaptive}, and epidemic networks \cite{huang2021temporal, deng2020colagnn}. The framework combines graph learning with transformer architectures to jointly capture temporal dynamics and cross-sectional dependencies.  A more recent, efficient
heterogeneous temporal GNN pursues the same goal with lower computational cost and explicitly positions
multiplex heterogeneous graph convolution as prior work in this space \cite{wang2025simple}. Memory-based temporal GNNs
(MTGNNs), which extend the TGN-style message/memory/embedding encoder-decoder design, have also
been adapted to multi-modal, multi-relational settings such as correlating financial news events with market movement \cite{Su2024}. 
Temporal heterogeneous GNNs have also been applied to the stock market.
\cite{xiang2022temporal} uses a temporal heterogeneous graph neural network (THGNN) for forecasting stock price movements by modelling both the temporal evolution of financial time series and the changing relationships among companies. The model constructs a daily graph from historical price data, combines transformer-based temporal feature extraction with a heterogeneous graph attention network, and predicts future price movements. Empirical results using U.S. and Chinese stock market data suggest the proposed approach consistently outperforms existing machine learning methods. To the best of our knowledge, our paper is the first to apply temporal multiplex to CDS forecasting, with an interpretable architecture that quantifies the time-varying contributions of distinct financial and liquidity interaction channels to credit-risk dynamics.

\section{Methodology}\label{sec3}
We model the global banking system as a temporal heterogeneous multiplex graph and develop a Temporal Heterogeneous Multiplex Graph Neural Network (Temporal HMGNN) to forecast bank-level credit risk and analyse systemic risk transmission. The proposed framework integrates bank-specific financial characteristics, country-level macroeconomic conditions, multiple channels of financial dependence, and temporal dependence within a unified architecture.

The modelling framework proceeds in five stages. First, bank- and country-level information is represented as node features. Second, three complementary network relations capture structural financial similarity, dynamic liquidity co-movement, and country-to-bank macroeconomic linkages. Third, a heterogeneous GNN encoder learns cross-sectional bank representations by processing the multiplex relations and combining their information. Fourth, a Gated Recurrent Unit (GRU) models the temporal evolution of the resulting bank embeddings. Finally, a nonlinear prediction layer maps the temporal representations into one-step-ahead changes in CDS spreads.
\subsection{Problem formulation}
Let \(\mathcal{B}=\{1,\ldots,N_B\}\) denote the set of $N_B$ bank nodes and
\(\mathcal{C}=\{1,\ldots,N_C\}\) the set of $N_C$ country nodes. At each forecast origin \(t\), the banking system is represented by

\[\mathcal{G}_{t}=\left(\mathcal{V},\mathcal{E}^{(F)},
\mathcal{E}_{t}^{(L)},\mathcal{E}^{(C\rightarrow B)},
\mathbf{X}_{t}^{(B)},\mathbf{X}_{t}^{(C)}\right),\]

\noindent where \(\mathcal{V}=\mathcal{B}\cup\mathcal{C}\) is the set of nodes, \(\mathcal{E}^{(F)}\) denotes the financial-similarity layer, \(\mathcal{E}_{t}^{(L)}\) the time-varying liquidity co-movement layer,
\(\mathcal{E}^{(C\rightarrow B)}\) the country-to-bank bipartite layer, \(\mathbf{X}_{t}^{(B)}\) is the bank feature matrix, and \(\mathbf{X}_{t}^{(C)}\) is the country feature matrix. Each bank \(i\) is associated with a log-transformed CDS spread $y_{i,t}=\log\!\left(1+\mathrm{CDS}_{i,t}\right)$. Using only information available up to time $t$, the first objective is to predict the one-step-ahead change
\[\Delta y_{i,t+1}=y_{i,t+1}-y_{i,t}\,.\]

\noindent Beyond prediction, our second objective is to learn the latent network structure governing financial transmission across institutions. The multiplex graph captures time-varying dependence relationships rather than direct contractual exposures, enabling the identification of contagion channels and the quantification of systemic importance.

\subsection{Temporal heterogeneous multiplex financial network}

\begin{figure}[H]
    \centering
    \includegraphics[width=\textwidth]{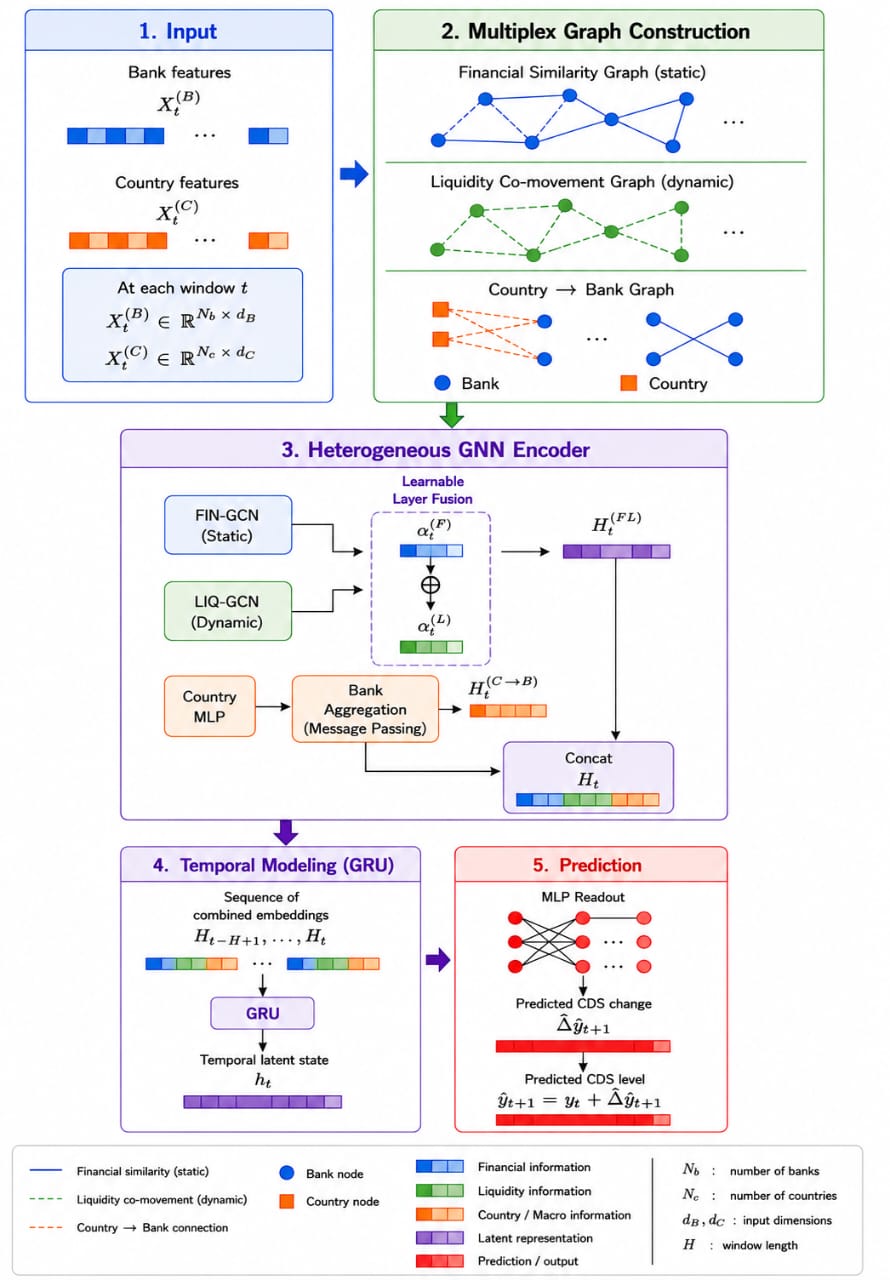}
    \caption{Model architecture}
    \label{thmgnn_architecture}
\end{figure}

Figure~\ref{thmgnn_architecture} presents an architectural overview of the proposed model. The model transforms a sequence of heterogeneous multiplex banking networks over a historical window $\tau=t-H+1,\ldots,t $ into a one-step-ahead CDS forecast. At each $\tau$, bank and country features are combined with three graph relations representing financial similarity, liquidity co-movement, and country-to-bank macroeconomic transmission. The heterogeneous GNN encoder first learns a cross-sectional representation of each bank. These representations are subsequently ordered through time and processed by a Gated Recurrent Unit (GRU) \citep{cho2014learning}, after which a nonlinear readout produces the predicted CDS change.

\subsubsection{Input features}
At each time period $\tau$, the proposed model considers two types of nodes: banks and countries. Let the feature vector of bank $i$ and country $c$ be denoted by
\[
\mathbf{x}^{(B)}_{i,\tau} \in \mathbb{R}^{d_B},
\qquad
\mathbf{x}^{(C)}_{c,\tau} \in \mathbb{R}^{d_C},
\]
respectively, where $d_B$ and $d_C$ denote the dimensions of the bank and country feature vectors. Collecting the node-level features gives
\[
\mathbf{X}^{(B)}_{\tau} \in \mathbb{R}^{N_B \times d_B},
\qquad
\mathbf{X}^{(C)}_{\tau} \in \mathbb{R}^{N_C \times d_C},
\]
where $N_B$ and $N_C$ denote the numbers of banks and countries, respectively.

The bank feature matrix contains institution-level information describing capital adequacy, liquidity, asset quality, balance-sheet structure, funding conditions, and lagged CDS information. Country nodes contain macroeconomic and financial-market indicators representing the economic environment associated with each bank. Section~\ref{sec4} describes the specific variables, data sources, transformations, and preprocessing procedures.

\noindent For each forecast origin $t$, the model uses a rolling historical window of length $H$, with $\tau=t-H+1,\ldots,t$.
The resulting input sequence is therefore
\[
\left\{
\mathbf{X}^{(B)}_{\tau},
\mathbf{X}^{(C)}_{\tau}
\right\}_{\tau=t-H+1}^{t},
\]
which provides the node-level information subsequently processed through the multiplex graph structure and heterogeneous GNN encoder.

\subsubsection{Multiplex graph construction}

\noindent To represent distinct channels of interconnectedness within the banking system, we construct a multiplex graph comprising two bank--bank layers and a heterogeneous country--to--bank layer. The three relations capture persistent financial similarity, time-varying liquidity co-movement, and macroeconomic transmission, respectively.

\noindent\textbf{Financial-similarity layer.}
Let $\mathbf{x}^{(F)}_{i,t}\in\mathbb{R}^{d_F}$ denote the vector of structural financial characteristics for bank $i$ at time $t$, where $d_F$ is the number of financial characteristics. For each quarter in the training period, the similarity between banks $i$ and $j$ is computed using an RBF kernel,
\[
w^{(F)}_{ij,t}
=
\exp\left(
-\frac{
\|\mathbf{x}^{(F)}_{i,t}-\mathbf{x}^{(F)}_{j,t}\|_2^2
}{
2\sigma_t^2
}
\right),
\]
where $\sigma_t$ is the median of the positive pairwise distances at time $t$. Let $\mathcal{T}_{\mathrm{train}}$ denote the set of quarters in the training period. The static financial adjacency matrix is obtained by averaging the quarter-specific similarities over the training period,
\[
A^{(F)}
=
\frac{1}{|\mathcal{T}_{\mathrm{train}}|}
\sum_{t\in\mathcal{T}_{\mathrm{train}}}
\left[w^{(F)}_{ij,t}\right]_{i,j=1}^{N_B}
\in\mathbb{R}^{N_B\times N_B}.
\]

\noindent To remove weak connections, top-$k$ sparsification is applied,
\[
\widetilde{A}^{(F)}
=
S_k\left(A^{(F)}\right),
\]
where $S_k(\cdot)$ retains the $k$ strongest off-diagonal connections for each bank, symmetrises the resulting neighbourhoods, removes edges below a minimum weight $\omega_{\min}$, and excludes self-loops. The resulting financial-similarity network is held fixed throughout forecasting and represents persistent structural similarities among banks.

\medskip
\noindent\textbf{Liquidity co-movement layer.}
Let $d_L$ denote the number of liquidity characteristics and
\[
\mathbf{z}_{i,t}
=
\left(
z^{(1)}_{i,t},\ldots,z^{(d_L)}_{i,t}
\right)^\top
\in\mathbb{R}^{d_L}
\]
the liquidity feature vector for bank $i$ at time $t$. For each bank, observations over the preceding $w$ quarters are concatenated to form the rolling liquidity-history vector
\[
\mathbf{q}_{i,t}
=
\left[
\left(z^{(1)}_{i,t-w+1},\ldots,z^{(1)}_{i,t}\right),
\ldots,
\left(z^{(d_L)}_{i,t-w+1},\ldots,z^{(d_L)}_{i,t}\right)
\right]^\top
\in\mathbb{R}^{wd_L}.
\]

\noindent Liquidity co-movement between banks $i$ and $j$ at time $t$ is measured by the correlation between their rolling liquidity histories, $\rho_{ij,t}=\operatorname{corr}
\left(\mathbf{q}_{i,t},\mathbf{q}_{j,t}\right)$. The correlation is transformed into a non-negative similarity weight,
\[
w^{(L)}_{ij,t}
=
\frac{1}{2}\left(1+\rho_{ij,t}\right),
\]
giving the time-varying liquidity adjacency matrix
\[
A^{(L)}_t
=
\left[w^{(L)}_{ij,t}\right]_{i,j=1}^{N_B}
\in\mathbb{R}^{N_B\times N_B}.
\]

\noindent As with the financial-similarity layer, top-$k$ sparsification is applied,
\[
\widetilde{A}^{(L)}_t
=
S_k\left(A^{(L)}_t\right).
\]
Unlike the financial-similarity layer, the liquidity network is recomputed at each quarter using rolling histories of the Liquidity Coverage Ratio (LCR), Loan-to-Deposit Ratio (LDR), and Deposit-to-Asset Ratio (DAR). It therefore captures evolving patterns of liquidity co-movement and funding dependence across banks.

\medskip
\noindent\textbf{Country--to--bank layer.}
Each bank is linked to its domicile country through a fixed bipartite graph. Let
\[
\pi:\mathcal{B}\rightarrow\mathcal{C}
\]
denote the mapping from banks to countries. The country--to--bank adjacency matrix is defined as
\[
A^{(C\rightarrow B)}_{ci}
=
\begin{cases}
1, & \pi(i)=c,\\
0, & \text{otherwise},
\end{cases}
\]
where $ A^{(C\rightarrow B)} \in \{0,1\}^{N_C\times N_B}$. Thus, each bank is connected to its corresponding country node, providing a heterogeneous transmission channel through which country-level macroeconomic information can subsequently be propagated to the bank during graph message passing.

At each time $\tau$, the resulting multiplex graph therefore consists of the static financial-similarity network $\widetilde{A}^{(F)}$, the dynamic liquidity co-movement network $\widetilde{A}^{(L)}_{\tau}$, and the fixed country--to--bank relation $A^{(C\rightarrow B)}$. Together, these layers represent persistent structural relationships, evolving liquidity dependencies, and macroeconomic transmission within a unified graph structure.

\subsubsection{Heterogeneous GNN encoder}

For each time period $\tau$, the heterogeneous GNN encoder transforms the bank and country node features, together with the multiplex graph structure, into a unified cross-sectional representation of the banking system. The encoder consists of an initial bank-feature projection, separate graph convolutions over the financial-similarity and liquidity co-movement layers, an adaptive fusion mechanism, and heterogeneous country--to--bank message passing.

\noindent\textbf{Initial bank representation.}
The raw bank features are first projected into a common hidden space. The initial bank representation at time $\tau$ is defined as
\[
H^{(0)}_{\tau}
=
\sigma\left(
X^{(B)}_{\tau}W_B+b_B
\right),
\]
where $W_B$ and $b_B$ are learnable parameters and $\sigma(\cdot)$ denotes the ReLU activation function. This transformation maps the bank-level input features into a $d$-dimensional latent space shared by the subsequent bank--bank graph convolutional layers.

\medskip
\noindent\textbf{Layer-wise bank--bank graph representation.}
The projected bank representations are propagated separately through the financial-similarity and liquidity co-movement networks. For the financial-similarity layer,
\[
H^{(F)}_{\tau}
=
\operatorname{GCN}_{F}
\left(
H^{(0)}_{\tau},
E^{(F)},
w^{(F)}
\right),
\]
where $\operatorname{GCN}_{F}(\cdot)$ (\cite{kipf2017semi}) denotes the weighted graph convolution applied to the static financial-similarity edge set $E^{(F)}$ with corresponding edge weights $w^{(F)}$.

Similarly, the liquidity-layer representation is
\[
H^{(L)}_{\tau}
=
\operatorname{GCN}_{L}
\left(
H^{(0)}_{\tau},
E^{(L)}_{\tau},
w^{(L)}_{\tau}
\right),
\]
where $E^{(L)}_{\tau}$ and $w^{(L)}_{\tau}$ denote the time-varying liquidity edge set and its corresponding edge weights.

Both $\operatorname{GCN}_{F}$ and $\operatorname{GCN}_{L}$ consist of stacks of weighted graph convolutional layers. Intermediate layers are followed by ReLU activation and dropout, while the final layer produces the relation-specific bank embeddings. Processing the two networks separately allows the model to learn distinct representations of persistent financial similarity and dynamic liquidity dependence.

\medskip
\noindent\textbf{Adaptive financial--liquidity multiplex fusion.}
The financial and liquidity embeddings are combined using a learnable, bank-specific, time-varying gating mechanism. For each bank $i$ at time $\tau$, the two relation-specific embeddings are concatenated and passed through a two-layer multilayer perceptron $g_{\theta}(\cdot)$:
\[s_{i,\tau}=g_{\theta}
\left(h_{i,\tau}^{(F)}\Vert h_{i,\tau}^{(L)}\right),\]

\noindent where $\Vert$ denotes vector concatenation and
$s_{i,\tau}\in\mathbb{R}^{2}$ contains the unnormalised financial and liquidity gating scores. The corresponding coefficients are obtained using a softmax transformation:

\[\left[\alpha_{i,\tau}^{(F)},\alpha_{i,\tau}^{(L)}\right]
=\operatorname{softmax}\left(\mathbf{s}_{i,\tau}\right),\]

\noindent where the softmax transformation is defined as

\[\alpha_{i,\tau}^{(k)}=\frac{\exp\!\left(s_{i,\tau}^{(k)}
\right)}{\sum\limits_{j\in\{F,L\}}
\exp\!\left(s_{i,\tau}^{(j)}\right)},\qquad k\in\{F,L\},\]

\noindent which satisfies
\[\alpha_{i,\tau}^{(F)}+\alpha_{i,\tau}^{(L)}=1,
\qquad \alpha_{i,\tau}^{(k)}\ge0.\]

\noindent The fused bank--bank representation is therefore
\[h_{i,\tau}^{(FL)}=\alpha_{i,\tau}^{(F)}
h_{i,\tau}^{(F)}+\alpha_{i,\tau}^{(L)}
h_{i,\tau}^{(L)}.\]

\noindent Equivalently, in matrix form,
\[H_{\tau}^{(FL)}=\alpha_{\tau}^{(F)}
\odot H_{\tau}^{(F)}+\alpha_{\tau}^{(L)}
\odot H_{\tau}^{(L)},\]

\noindent where $\odot$ denotes row-wise multiplication through broadcasting. The coefficients are bank-specific and time-varying because they depend on each bank's financial and liquidity embeddings at each time period. Larger
values of $\alpha_{i,\tau}^{(F)}$ indicate a greater contribution from persistent balance-sheet similarity, whereas larger values of $\alpha_{i,\tau}^{(L)}$ indicate that liquidity co-movement is more influential
for the representation of bank $i$ at time $\tau$.

Importantly, the gating mechanism operates only over the financial and liquidity bank-bank layers. Country information is incorporated separately through the heterogeneous country-to-bank relation.

\medskip
\noindent\textbf{Country--to--bank macroeconomic propagation.}
Country features are encoded separately using a multilayer perceptron,
\[
H^{(C)}_{\tau}
=
\phi_C
\left(
X^{(C)}_{\tau}
\right),
\]
where $\phi_C(\cdot)$ consists of two linear transformations with an intermediate ReLU activation.

 The resulting country embeddings are propagated to the associated bank nodes through the country-to-bank bipartite network using GraphSAGE (\cite{hamilton2017inductive}), an inductive graph representation method that aggregates information from neighbouring nodes:
\[
\widetilde{H}^{(C\rightarrow B)}_{\tau}
=
\operatorname{SAGE}_{C\rightarrow B}
\left(
H^{(C)}_{\tau},
H^{(FL)}_{\tau},
E^{(C\rightarrow B)}
\right).
\]
A ReLU activation is subsequently applied,
\[
H^{(C\rightarrow B)}_{\tau}
=
\sigma
\left(
\widetilde{H}^{(C\rightarrow B)}_{\tau}
\right).
\]

\noindent Here, country nodes act as source nodes and bank nodes as destination nodes, with $E^{(C\rightarrow B)}$ linking each bank to its domicile country. The GraphSAGE operation combines the relevant country embedding with the destination bank's existing financial-liquidity representation, allowing country-level macroeconomic conditions to affect bank representations while preserving heterogeneity between bank and country nodes.

\medskip
\noindent\textbf{Final cross-sectional bank representation.}
The fused bank--bank representation and the country-informed bank representation are combined using fixed equal weights:
\[
H_{\tau}
=
\frac{1}{2}H^{(FL)}_{\tau}
+
\frac{1}{2}H^{(C\rightarrow B)}_{\tau}.
\]

\noindent The resulting matrix $H_{\tau}\in\mathbb{R}^{N_B\times d}$ 
contains the final cross-sectional representation of all banks at time $\tau$. Thus, the adaptive weights are learned specifically for the financial and liquidity bank--bank channels, whereas the subsequent integration of country-level information uses fixed coefficients. Repeating this encoding procedure over the historical window produces a sequence of bank representations that is subsequently passed to the temporal modelling component.

\subsubsection{Temporal modelling using GRU}

\noindent The heterogeneous GNN encoder produces a cross-sectional bank representation $H_{\tau}$ at each time period $\tau$. To capture the evolution of bank credit risk and network dependencies over time, the representations obtained over a rolling historical window of length $H$ are arranged chronologically as
\[
\left\{
H_{t-H+1}, H_{t-H+2}, \ldots, H_t
\right\}.
\]

\noindent The resulting sequence is passed through a Gated Recurrent Unit (GRU) \cite{cho2014learning}:
\[
Z_{t-H+1:t}
=
\operatorname{GRU}
\left(
H_{t-H+1}, H_{t-H+2}, \ldots, H_t
\right).
\]

\noindent The GRU processes the sequence of graph-informed representations for each bank and produces a temporal representation at each point in the historical window. For one-step-ahead forecasting, only the final GRU output is retained:
\[
Z_t
=
Z_{t-H+1:t}[-1], \quad \text{where} \quad Z_t \in \mathbb{R}^{N_B\times d}\,.
\]

\noindent For bank $i$, the corresponding row $z_{i,t}$ therefore summarises its graph-informed history up to forecast origin $t$. In this way, the temporal component complements the heterogeneous GNN encoder by capturing persistence in credit risk, delayed transmission effects, and the evolution of network dependencies across the historical window. The resulting temporal representation $Z_t$ is subsequently passed to the prediction layer to generate the one-step-ahead CDS forecast.

\subsubsection{Prediction and learning objective}

\noindent The final temporal representation obtained from the GRU is used to generate the one-step-ahead CDS forecast. For each bank $i$, let $z_{i,t}$ denote the $i$th row of $Z_t$. The predicted change in the log-transformed CDS spread is obtained through a nonlinear readout network,
\[
\widehat{\Delta y}_{i,t+1}
=
f_{\mathrm{readout}}(z_{i,t}),
\]
where $f_{\mathrm{readout}}(\cdot)$ is a multilayer perceptron with an intermediate ReLU activation. The forecasting target is defined as
\[
\Delta y_{i,t+1}
=
y_{i,t+1}-y_{i,t}
=
\log(1+\mathrm{CDS}_{i,t+1})
-
\log(1+\mathrm{CDS}_{i,t}).
\]

\noindent The predicted log-CDS level can subsequently be reconstructed from the predicted change as
\[\widehat{y}_{i,t+1}=y_{i,t}+\widehat{\Delta y}_{i,t+1}.\]

\noindent Hence, the complete architecture combines two weighted bank-bank graph convolutional channels, an adaptive financial-liquidity fusion gate, a heterogeneous country-to-bank GraphSAGE layer, and a GRU-based temporal aggregator. While existing financial temporal GNNs model cross-market dependencies and their temporal evolution for forecasting (e.g., \cite{son2023forecasting}), the proposed architecture additionally integrates multiple bank-interaction channels, heterogeneous bank and country nodes, adaptive financial-liquidity fusion, and time-varying network dependencies. The learned financial-liquidity gating coefficients also provide
a direct indication of whether structural balance-sheet similarity or liquidity co-movement contributes more strongly to the bank representation under different market conditions.\\

\noindent Furthermore, let $ \Theta=\left\{\Theta_{B},\Theta_{F},\Theta_{L},\Theta_{\mathrm{gate}},
\Theta_{C},\Theta_{C\rightarrow B},\Theta_{\mathrm{GRU}},
\Theta_{\mathrm{readout}}\right\}$ denote the complete set of trainable model parameters. Here, $\Theta_{B}$ contains the parameters of the initial bank projection; $\Theta_{F}$ and $\Theta_{L}$ contain the parameters of the financial and
liquidity GCN stacks; $\Theta_{\mathrm{gate}}$ contains the parameters of the financial--liquidity gating network; $\Theta_{C}$ contains the country encoder parameters; $\Theta_{C\rightarrow B}$ contains the country-to-bank GraphSAGE parameters; and $\Theta_{\mathrm{GRU}}$ and
$\Theta_{\mathrm{readout}}$ contain the temporal and prediction-layer parameters, respectively. The proposed Temporal HMGNN can then be expressed compactly as
\[
\widehat{\Delta y}_{i,t+1}
=
f_{\Theta}
\left(
X^{(B)}_{t-H+1:t},
X^{(C)}_{t-H+1:t},
E^{(F)},w^{(F)},
E^{(L)}_{t-H+1:t},w^{(L)}_{t-H+1:t},
E^{(C\rightarrow B)}
\right),
\]
where $f_{\Theta}(\cdot)$ represents the complete mapping from the historical sequence of node features and multiplex graph structures to the one-step-ahead CDS change. The model parameters are estimated by minimising the Huber loss over the set $\mathcal{D}$ of bank-time observations for which the next-period CDS target is available:
\[
\Theta^{*}
=
\arg\min_{\Theta}\mathcal{L}(\Theta),
\]
where
\[
\mathcal{L}(\Theta)
=
\frac{1}{|\mathcal{D}|}
\sum_{(i,t)\in\mathcal{D}}
\ell_{\delta}
\left(
\widehat{\Delta y}_{i,t+1}
-
\Delta y_{i,t+1}
\right).
\]

\noindent For a forecast error $r_{i,t+1}
=\widehat{\Delta y}_{i,t+1}-\Delta y_{i,t+1}$, the Huber loss is defined as
\[
\ell_{\delta}(r)
=
\begin{cases}
\dfrac{1}{2}r^2, & |r|\leq\delta,\\[6pt]
\delta\left(|r|-\dfrac{1}{2}\delta\right), & |r|>\delta.
\end{cases}
\]
Consistent with the model implementation, we set the threshold parameter to $\delta=0.5$. The Huber loss is quadratic for small forecast errors and linear for large errors, thereby reducing the influence of extreme CDS movements while retaining sensitivity to deviations around the centre of the error distribution. Model parameters are optimised using Adam with a learning rate of $10^{-3}$ and weight decay of $10^{-4}$. At each forecast origin $t$, the model uses only bank features, country features, CDS information, and graph structures available up to time $t$, thereby preventing look-ahead bias.

Algorithm~\ref{alg:temporal_hgnn} summarises the complete procedure. It first constructs the static financial-similarity network, the dynamic liquidity co-movement networks, and the country-to-bank bipartite network. At each time period, the model learns separate financial and liquidity bank embeddings and combines them using a bank-specific softmax gate. The model then encodes country features and propagates them to banks through the heterogeneous GraphSAGE layer. The financial-liquidity and country-informed bank representations are averaged to obtain the final cross-sectional embedding. The resulting sequence is processed by the GRU, and its final temporal state is used to predict the one-step-ahead change in log CDS spread.

\begin{algorithm}[H]
\caption{Temporal HMGNN for CDS Forecasting}
\label{alg:temporal_hgnn}
\begin{algorithmic}[1]

\Require Bank features $\mathbf{X}^{(B)}_{\tau}$, country features $\mathbf{X}^{(C)}_{\tau}$, CDS spreads $CDS_{i,\tau}$, history length $H$
\Ensure Predicted CDS change $\widehat{\Delta y}_{i,t+1}$

\State \textbf{Target construction:}
\State $y_{i,t} \gets \log(1+CDS_{i,t})$
\State $\Delta y_{i,t+1} \gets y_{i,t+1}-y_{i,t}$

\State \textbf{Graph construction:}
\State Construct the static financial-similarity layer
$\mathcal{E}^{(F)}$ with edge weights $\mathbf{w}^{(F)}$
from balance-sheet features using top-$k$ sparsification
\State Construct the country-to-bank bipartite layer
$\mathcal{E}^{(C\rightarrow B)}$
linking each bank to its domicile country
\For{each time period $\tau$}
    \State Construct the dynamic liquidity co-movement layer
    $\mathcal{E}_{\tau}^{(L)}$ with edge weights
    $\mathbf{w}_{\tau}^{(L)}$
    from rolling correlations of liquidity variables
\EndFor

\State \textbf{Temporal multiplex encoding:}
\For{$\tau=t-H+1$ to $t$}

    \State $\mathbf{H}_{\tau}^{(0)}
    \gets
    \sigma\!\left(
    \mathbf{X}_{\tau}^{(B)}\mathbf{W}_{B}
    +\mathbf{b}_{B}
    \right)$

    \State $\mathbf{H}_{\tau}^{(F)}
    \gets
    \operatorname{GCN}_{F}
    \left(
    \mathbf{H}_{\tau}^{(0)},
    \mathcal{E}^{(F)},
    \mathbf{w}^{(F)}
    \right)$

    \State $\mathbf{H}_{\tau}^{(L)}
    \gets
    \operatorname{GCN}_{L}
    \left(
    \mathbf{H}_{\tau}^{(0)},
    \mathcal{E}_{\tau}^{(L)},
    \mathbf{w}_{\tau}^{(L)}
    \right)$

    \For{each bank $i$}
        \State $\mathbf{s}_{i,\tau}
        \gets
        g_{\theta}
        \left(
        \mathbf{h}_{i,\tau}^{(F)}
        \Vert
        \mathbf{h}_{i,\tau}^{(L)}
        \right)$

        \State $\left[
        \alpha_{i,\tau}^{(F)},
        \alpha_{i,\tau}^{(L)}
        \right]
        \gets
        \operatorname{softmax}
        \left(
        \mathbf{s}_{i,\tau}
        \right)$

        \State $\mathbf{h}_{i,\tau}^{(FL)}
        \gets
        \alpha_{i,\tau}^{(F)}
        \mathbf{h}_{i,\tau}^{(F)}
        +
        \alpha_{i,\tau}^{(L)}
        \mathbf{h}_{i,\tau}^{(L)}$
    \EndFor

    \State $\mathbf{H}_{\tau}^{(C)}
    \gets
    \phi_{C}
    \left(
    \mathbf{X}_{\tau}^{(C)}
    \right)$

    \State $\mathbf{H}_{\tau}^{(C\rightarrow B)}
    \gets
    \sigma\!\left(
    \operatorname{SAGE}_{C\rightarrow B}
    \left(
    \mathbf{H}_{\tau}^{(C)},
    \mathbf{H}_{\tau}^{(FL)},
    \mathcal{E}^{(C\rightarrow B)}
    \right)
    \right)$

    \State $\mathbf{H}_{\tau}
    \gets
    \frac{1}{2}\mathbf{H}_{\tau}^{(FL)}
    +
    \frac{1}{2}\mathbf{H}_{\tau}^{(C\rightarrow B)}$

\EndFor

\State \textbf{Temporal aggregation:}
\State $\mathbf{Z}_{t}
\gets
\operatorname{GRU}
\left(
\mathbf{H}_{t-H+1},
\ldots,
\mathbf{H}_{t}
\right)_{\mathrm{last}}$

\State \textbf{Prediction:}
\For{each bank $i$}
    \State $\widehat{\Delta y}_{i,t+1}
    \gets
    f\left(\mathbf{z}_{i,t}\right)$
\EndFor

\State \textbf{Training:}
\State Minimise the Huber loss over bank--time observations with available CDS targets:
\Statex \hspace{\algorithmicindent}
$\displaystyle
\mathcal{L}(\Theta)
=
\frac{1}{|\mathcal{D}|}
\sum_{(i,t)\in\mathcal{D}}
\ell_{\delta}
\left(
\widehat{\Delta y}_{i,t+1}
-
\Delta y_{i,t+1}
\right)
$

\State Use only information available up to time $t$ to prevent look-ahead bias

\State \Return $\widehat{\Delta y}_{i,t+1}$

\end{algorithmic}
\end{algorithm}

\section{Data and Empirical Design}\label{sec4}

This section describes the data and empirical framework used to evaluate the proposed Temporal HMGNN. We first present the bank-level, CDS, and macroeconomic data and their preprocessing, followed by the experimental design, benchmark models, and evaluation metrics used for out-of-sample assessment.

\subsection{Data and variable construction}
\subsubsection{Bank sample and bank-level fundamentals}
The empirical analysis is conducted on a panel of 29 large banks, comprising global SIBs and major banks spanning North America, Europe, and Asia-Pacific. The sample includes major banking institutions such as JPMorgan Chase, HSBC, BNP Paribas, and Mitsubishi UFJ, thereby ensuring that key nodes within the global financial system are adequately covered. The full list of institutions and their acronyms used throughout the analysis is reported in the Appendix Table~\ref{tab:bank_list}. Bank-level fundamentals (1998Q1 to 2025Q4) are obtained from a structured dataset of financial statements and regulatory disclosures available from \emph{Bloomberg}. The raw data are reorganised into a quarterly panel format, where each observation corresponds to a bank-quarter pair and the following core variables are retained: 

Capital adequacy - \emph{CET1 ratio, Tier 1 ratio}, Liquidity - \emph{LCR}, Asset quality - \emph{Non-performing loan (NPL) ratio}, Balance sheet scale - \emph{Total assets, Risk-weighted assets (RWA)}, Funding structure -  \emph{Deposits, Loans}. In addition, to capture risk-taking behaviour, structural characteristics and regulatory capital intensity, we further defined the following financial ratios:
\emph{LDR = $\frac{Loans}{Deposits}$}, \emph{DAR = $\frac{Deposits}{Total \;Assets}$} and \emph{RWA Density = $\frac{RWA}{Total \; Assets}$.}

\subsubsection{CDS data}
Credit risk is proxied by 5-year CDS spreads, collected daily and aggregated to quarterly frequency via cross-sectional averaging. Due to data availability constraints, CDS observations are available for a subset of banks (17 out of 29), from 2001Q3 onwards, thereby yielding 1249 bank-quarter CDS observations. Importantly, we retain all banks in the network structure to preserve systemic relationships, while evaluating the forecasting task only on observations with valid CDS targets. This approach ensures that the model leverages the full cross-sectional network information without introducing bias from missing target variables.

\subsubsection{Macroeconomic variables}
Macroeconomic conditions are incorporated at the country level, reflecting the economic environment in which each bank operates. Two primary data sources are used:
\begin{itemize}
    \item Global Financial Indicators (FRED): VIX (market volatility), US high-yield spread, US 10-year Treasury yield, Federal Funds rate.
\item Country-Level Macroeconomic Indicators (World Bank): GDP growth (YoY), 
Inflation (CPI), Unemployment rate, Lending rate, Credit to private sector (\% GDP).
\end{itemize}
Annual World Bank data are converted to a quarterly frequency by replicating values across quarters within each year, then forward-filled to match the panel horizon. Finally, each bank is mapped to its country of domicile, and macroeconomic variables are merged accordingly, creating a hierarchical structure linking country conditions to bank-level risk.

\subsubsection{Node Features}
The model operates on a temporal heterogeneous graph, where each node represents either a bank or a country. We define the bank node and country node features as follows:\\

\noindent \textbf{Bank Node Features}: At each time $t$, the bank nodes are described by financial ratios (capital, liquidity, asset quality), balance sheet variables, macroeconomic variables (country-linked) and lagged CDS features. To strictly avoid data leakage, we standardised using training-set statistics only and imputed using cross-sectional medians where necessary.\\

\noindent \textbf{Country Node Features}: Country nodes are represented by macroeconomic indicators (World Bank + FRED) and are standardised using training-period scaling.\\

\noindent \textbf{Graph Structure}: Our model incorporates three types of relationships:
\begin{itemize}
\item Static financial similarity network (bank–bank): This network is based on balance sheet characteristics. We construct it using RBF similarity on financial variables and then average it over the training period.
\item Dynamic liquidity network (bank–bank): Here, we consider a time-varying network based on rolling correlations of liquidity metrics (LCR, LDR, DAR) to capture evolving interbank dependencies.
\item Country-to-bank links (heterogeneous edges): Here, each bank is connected to its country node, thereby enabling macro-to-micro information transmission.
\end{itemize}

\subsubsection{Final panel construction}

The final dataset is constructed as a balanced quarterly panel by integrating bank-level fundamentals, CDS spreads, and country-level macroeconomic variables into a unified framework. All variables are aligned to quarter-end timestamps to ensure temporal consistency across sources. Missing observations are handled using a combination of forward filling (restricted to a maximum of two consecutive quarters) and cross-sectional median imputation to preserve data integrity while avoiding excessive smoothing. From this processed panel, we construct time-indexed bank node feature matrices $X_t$, country feature matrices $C_t$, and forecasting targets $\Delta y_{i,t}$, which serve as inputs to the temporal graph learning framework. 

The resulting dataset spans the period 1998–2025, encompassing major financial episodes such as the dot-com crash (2000–2002), the Global Financial Crisis (2007–2009), the European sovereign debt crisis (2010–2012), and the COVID-19 shock (2020–2021). This extensive time horizon allows the model to capture not only long-term structural relationships but also cyclical fluctuations and crisis-driven dynamics within the global banking network. Overall, the dataset provides a rich, multi-layered representation of global banking risk, combining institution-level fundamentals, market-based credit risk measures, and macroeconomic conditions within a unified temporal network framework.

To further characterise the empirical properties of the final panel, Table~\ref{tab:summary_stats} reports summary statistics for the bank-level and macroeconomic variables used in the analysis. The bank-level variables exhibit substantial cross-sectional heterogeneity in capital adequacy, liquidity conditions, balance-sheet structure, and market-implied credit risk. In particular, CDS spreads display considerable dispersion, reflecting pronounced variation in perceived credit risk across institutions and over time. Similarly, liquidity and funding-related measures, such as the loan-to-deposit ratio and deposit-to-asset ratio, exhibit notable variability, supporting the importance of modelling heterogeneous and time-varying dependence structures across banks. The macroeconomic variables also capture diverse economic and financial conditions across countries and periods, including episodes of elevated market volatility and financial stress.

\begin{table}[H]
\centering
\caption{Summary statistics of variables used in the empirical analysis.}
\label{tab:summary_stats}
\small
\begin{tabular}{lrrrrrr}
\toprule
Variable & Obs. & Mean & Std. Dev. & Median & Min & Max \\
\midrule

\multicolumn{7}{l}{\textbf{Panel A: Bank-Level Variables}} \\
\midrule

CET1 Ratio & 1418 & 12.658 & 2.255 & 12.600 & 2.160 & 22.400 \\
Tier 1 Ratio & 2330 & 12.796 & 3.468 & 12.985 & 5.450 & 29.300 \\
Liquidity Coverage Ratio & 1184 & 136.955 & 21.395 & 134.000 & 67.520 & 324.900 \\
NPL Ratio & 1939 & 2.261 & 2.273 & 1.575 & 0.035 & 17.459 \\
RWA Density & 2122 & 0.415 & 0.171 & 0.391 & 0.035 & 3.557 \\
Loan-to-Deposit Ratio & 2346 & 0.868 & 0.347 & 0.769 & 0.278 & 2.394 \\
Deposit-to-Asset Ratio & 2431 & 0.516 & 0.165 & 0.535 & 0.139 & 0.862 \\
Log Total Assets & 2458 & 13.936 & 0.924 & 14.178 & 9.204 & 15.819 \\
5Y CDS Spread & 1249 & 76.231 & 65.603 & 60.119 & 4.765 & 500.020 \\
Log CDS Spread & 1249 & 4.046 & 0.803 & 4.113 & 1.752 & 6.217 \\

\midrule
\multicolumn{7}{l}{\textbf{Panel B: Macroeconomic and Market Variables}} \\
\midrule

GDP Growth & 1680 & 2.327 & 3.049 & 2.135 & -10.940 & 14.520 \\
CPI Inflation & 1680 & 1.927 & 1.657 & 1.796 & -1.401 & 10.001 \\
Unemployment Rate & 1680 & 6.472 & 3.418 & 5.407 & 2.119 & 26.094 \\
Lending Rate & 1076 & 3.996 & 2.064 & 4.350 & 0.500 & 9.233 \\
Private Credit-to-GDP & 1608 & 125.122 & 39.720 & 122.952 & 34.096 & 223.842 \\
VIX & 1680 & 20.206 & 7.272 & 18.507 & 10.310 & 58.753 \\
US High-Yield Spread & 1680 & 5.302 & 2.484 & 4.660 & 2.640 & 17.737 \\
US 10-Year Yield & 1680 & 3.474 & 1.355 & 3.607 & 0.650 & 6.480 \\
Federal Funds Rate & 1680 & 2.233 & 2.115 & 1.545 & 0.060 & 6.520 \\

\bottomrule
\end{tabular}
\end{table}

The summary statistics indicate substantial heterogeneity across institutions in terms of capital adequacy, liquidity conditions, funding structure, and market-implied credit risk. In particular, CDS spreads exhibit considerable dispersion, reflecting large differences in perceived default risk across banks and periods, especially during periods of financial stress. Liquidity and balance-sheet variables also display notable variability, supporting the importance of modelling multiple channels of interconnectedness within the banking system. Furthermore, the macroeconomic variables capture diverse economic and financial conditions, including periods of elevated market volatility and severe economic contraction. Overall, the results highlight the complex, nonlinear, and time-varying nature of systemic banking risk, motivating a temporal multiplex graph learning framework.

\subsection{Empirical Design}

We evaluate and compare the performance of the proposed Temporal HMGNN through out-of-sample forecasting against a diverse set of benchmark models. We then evaluate forecast performance using complementary accuracy, statistical significance, and directional measures.

\subsubsection{Benchmark models}
 
To assess the incremental value of incorporating multiplex interactions, heterogeneous node types, and temporal dependencies in the proposed Temporal HMGNN, we compare its predictive performance with benchmark models spanning econometric, machine-learning, sequence-based, and graph-based approaches. The benchmarks are grouped into four categories: 
\begin{itemize}
\item \textit{Persistence baseline}: Last Value; 
\item \textit{Econometric models}: Panel Fixed Effects and Bayesian Vector Autoregression (BVAR); 
\item \textit{Non-graph machine-learning models}: XGBoost and Long Short-Term Memory (LSTM); and 
\item \textit{Graph-based models}: Temporal GCN and Temporal Graph Attention Network (TGAT).
\end{itemize}

The Last Value model provides a persistence benchmark in which the next-period log-CDS spread is predicted by its most recent observed value. The Panel Fixed Effects model provides a linear benchmark that accounts for bank-specific heterogeneity, while the BVAR captures linear cross-bank dependencies through lagged CDS dynamics. XGBoost provides a nonlinear machine-learning benchmark based on lagged bank-level information, whereas LSTM captures nonlinear temporal dependencies through recurrent sequence modelling.

For the graph-based benchmarks, TGCN combines graph convolution \citep{kipf2017semi} with GRU-based temporal modelling \citep{cho2014learning}, while TGAT replaces the graph convolution with a graph attention mechanism \citep{velivckovic2017graph}. Both models use the time-varying liquidity co-movement network as the underlying bank-bank graph and therefore capture dynamic cross-sectional and temporal dependencies. However, unlike the proposed Temporal HMGNN, they operate on a single bank-bank relation and do not incorporate the financial-similarity layer, heterogeneous country nodes, or country-to-bank macroeconomic transmission.

For comparability, the machine-learning, sequence, and graph-based benchmarks use the same historical window and one-step-ahead forecasting horizon as the proposed model. The benchmark set therefore provides progressively richer representations of the forecasting problem, ranging from simple persistence and linear dependence to nonlinear temporal and graph-based dependence. This allows us to assess the incremental value of the proposed architecture in terms of its joint modelling of multiplex bank interactions, heterogeneous bank--country relationships, and temporal dynamics.

\subsubsection{Evaluation Metrics}

We evaluate model performance out of sample using complementary measures of forecast accuracy, statistical significance, and directional performance. Point forecast accuracy is assessed using Mean Squared Error (MSE), Mean Absolute Error (MAE), the coefficient of determination ($R^2$), and Pearson correlation. We evaluate statistical differences in predictive accuracy between the proposed Temporal HMGNN and the competing benchmark models using the Diebold-Mariano (DM) test \citep{diebold2002comparing}. Directional forecasting performance is assessed using directional accuracy (DA), together with a binomial test against the 50\% random-guessing benchmark and the Pesaran-Timmermann (PT) test \citep{pesaran1992simple} for statistical dependence between predicted and realised CDS directions.

\section{Empirical Results and discussion}\label{sec5}
\subsection{Predictive Performance}
Table~\ref{tab:point} reports point forecast accuracy across all models using standard error-based and goodness-of-fit metrics.
\begin{table}[H]
\centering
\caption{Point Forecast Accuracy}
\label{tab:point}
\begin{tabular}{lcccc}
\toprule
Model & MSE & MAE & $R^2$ & Corr \\
\midrule
Temporal HMGNN & \textbf{0.0309} & \textbf{0.1342} & \textbf{0.8136} & \textbf{0.9032} \\
Baseline (Last Value) & 0.0328 & 0.1425 & 0.8021 & 0.9008 \\
Temporal GAT & 0.0338 & 0.1403 & 0.7961 & 0.9022 \\
Temporal GCN & 0.0353 & 0.1463 & 0.7871 & 0.8923 \\
XGBoost & 0.0364 & 0.1573 & 0.7803 & 0.8994 \\
LSTM & 0.0447 & 0.1729 & 0.7304 & 0.8970 \\
BVAR & 0.0489 & 0.1736 & 0.7096 & 0.8508 \\
Panel Regression & 0.0596 & 0.1958 & 0.6404 & 0.8900 \\
\bottomrule
\end{tabular}
\end{table}
The proposed Temporal HMGNN achieves the strongest overall performance across all evaluation metrics. It records the lowest mean squared error (MSE = 0.0309) and mean absolute error (MAE = 0.1342), indicating superior accuracy in predicting CDS spread levels. In addition, it attains the highest coefficient of determination ($R^2 = 0.8136$) and correlation (0.9032), demonstrating strong explanatory power and close alignment with observed values. Among benchmark models, the naive baseline (last value) performs competitively, reflecting the persistence typically observed in CDS spreads. The strong performance of the naive baseline further highlights the persistence and autocorrelation inherent in CDS spreads, underscoring the challenge of achieving substantial improvements over simple benchmarks. Temporal graph-based models (GAT and GCN) also perform well but remain consistently below the Temporal HMGNN. 

In contrast, traditional machine learning and econometric models, including XGBoost, LSTM, BVAR, and panel regression, exhibit higher forecast errors and lower explanatory power. Overall, these results indicate that incorporating multiplex network structure and temporal dependencies substantially improves point forecast accuracy.

\subsection{Directional accuracy and Statistical validation}
\subsubsection{Directional accuracy}
Table~\ref{DForecast} reports directional forecasting performance across all models, including DA, binomial test p-values, and PT statistics. These metrics jointly assess both the magnitude and the statistical significance of sign-prediction performance.

\begin{table}[H]
\centering
\caption{Directional Forecast Evaluation (Bank-Level CDS Changes)}
\label{DForecast}
\begin{tabular}{lccccccc}
\toprule
Model & DA & Correct & $N$ & Binomial $p$ & PT stat & PT $p$ & $P^*$ \\
\midrule
XGBoost & 0.5781 & 185 & 320 & 0.0030 & 2.7602 & 0.0058 & 0.5010 \\
Temporal HMGNN & 0.5406 & 173 & 320 & 0.0811 & 1.4604 & 0.1442 & 0.4998 \\
LSTM & 0.5313 & 170 & 320 & 0.1441 & 1.0761 & 0.2819 & 0.5012 \\
Panel Regression & 0.5313 & 170 & 320 & 0.1441 & 1.2019 & 0.2294 & 0.4977 \\
Temporal GAT & 0.5281 & 169 & 320 & 0.1710 & 0.9433 & 0.3455 & 0.5018 \\
Temporal GCN & 0.5250 & 168 & 320 & 0.2009 & 0.9923 & 0.3211 & 0.4973 \\
Baseline (Last Value) & 0.5188 & 166 & 320 & 0.2693 & 0.6289 & 0.5294 & 0.5012 \\
BVAR & 0.4811 & 127 & 264 & 0.7508 & -0.5987 & 0.5494 & 0.4995 \\
\bottomrule
\end{tabular}
\end{table}

{\footnotesize \noindent DA denotes the proportion of correctly predicted CDS spread changes. Binomial $p$-values test whether DA differs from 50\% under random guessing. The PT test evaluates the statistical independence between predicted and realised directions. $P^*$ denotes the expected success probability under independence.}

Among all models, XGBoost achieves the highest directional accuracy (57.81\%) and is the only model to show statistically significant directional predictive ability under both the binomial test ($p < 0.01$) and the PT test ($p < 0.01$). This indicates that its directional forecasts are significantly better than random chance. The Temporal HMGNN achieves a directional accuracy of 54.06\%, outperforming several benchmark models by a large margin. However, its performance is not statistically significant under either the binomial test ($p = 0.081$) or the PT test ($p = 0.144$), suggesting that its directional predictions are not reliably distinguishable from random variation. All remaining models, including LSTM, panel regression, and temporal graph-based architectures (GCN and GAT), exhibit directional accuracies close to 50\% and fail to achieve statistical significance. 

In contrast, while XGBoost demonstrates superior performance in directional forecasting, it operates as a feature-driven model that does not explicitly account for interbank network structure. In contrast, the proposed Temporal HMGNN is designed to capture cross-sectional dependencies, contagion channels, and systemic risk propagation across financial institutions. The absence of statistically significant directional accuracy for the Temporal HMGNN does not undermine its primary contribution. Rather, it reflects that the model is optimised to learn structural and temporal interactions within the financial network, not short-term sign prediction. 

This highlights a fundamental distinction: traditional machine learning models may excel in isolated predictive tasks, whereas graph-based frameworks provide a richer representation of systemic interdependencies, enabling analyses of contagion dynamics, systemic importance, and macro-financial transmission mechanisms that standard approaches cannot capture.

\subsubsection{Statistical validation}
Next, we conduct the pairwise Diebold–Mariano (DM) tests between the Temporal HMGNN and each benchmark model to assess whether differences in predictive accuracy are statistically significant. The test is based on the difference in squared forecast errors and accounts for serial correlation in forecast residuals. Table~\ref{tab:dm} reports the DM test results comparing the predictive accuracy of the proposed Temporal HMGNN (Model 1) against benchmark models (Model 2). The DM test is based on forecast error differentials (typically squared errors) and therefore evaluates differences in magnitude prediction accuracy rather than directional accuracy.

\begin{table}[H]
\centering
\caption{Diebold--Mariano Test: Temporal HMGNN vs Benchmark Models}
\label{tab:dm}
\begin{tabular}{lcccc}
\toprule
Model 2 & $N$ & DM Statistic & $p$-value & Interpretation \\
\midrule
Panel Regression & 336 & -6.5222 & 6.93e-11 & HMGNN significantly more accurate \\
LSTM Baseline & 336 & -4.5960 & 4.31e-06 & HMGNN significantly more accurate \\
BVAR & 263 & -3.1483 & 0.0016 & HMGNN significantly more accurate \\
XGBoost & 336 & -2.9367 & 0.0033 & HMGNN significantly more accurate \\
Temporal GAT & 336 & -2.7119 & 0.0067 & HMGNN significantly more accurate \\
Temporal GCN & 336 & -2.6976 & 0.0070 & HMGNN significantly more accurate \\
Baseline (Last Value) & 336 & -1.0151 & 0.3100 & No significant difference \\
\bottomrule
\end{tabular}
\end{table}

The results in Table~\ref{tab:dm} indicate that the Temporal HMGNN consistently outperforms all benchmark models in terms of predictive accuracy. The DM statistics are negative across all comparisons, indicating lower forecast error for the Temporal HMGNN than for competing models. This improvement is statistically significant at conventional levels for all models except the naive baseline. In particular, the Temporal HMGNN significantly outperforms Panel Regression, LSTM, BVAR, XGBoost, and temporal graph-based models (GCN and GAT), with p-values well below 1\%. No statistically significant difference is observed relative to the baseline (last value), suggesting that naive persistence remains difficult to outperform in certain periods.

These findings provide strong evidence that the proposed Temporal HMGNN delivers superior point forecast accuracy relative to both traditional econometric models and machine learning benchmarks. Importantly, this contrasts with the directional forecasting results, in which XGBoost exhibits stronger sign-prediction performance. This highlights a key distinction: while feature-based models may capture short-term directional movements, the Temporal HMGNN is more effective at modelling the magnitude and dynamics of CDS spreads by incorporating interbank network structure and temporal dependencies.

Overall, the DM test results reinforce the Temporal HMGNN framework as a robust tool for modelling systemic risk and cross-sectional dependencies in financial networks.

\subsection{Interpretability Design}

Beyond predictive performance, a key contribution of the proposed framework lies in its ability to provide structural insights into systemic risk and contagion dynamics within the global banking network. This interpretability framework enables the analysis of systemic importance (which institutions matter most), contagion pathways (how shocks propagate), and macro-financial transmission (the role of economic conditions).

While all models are evaluated using the same predictive metrics to ensure a fair comparison, their interpretability differs substantially in both scope and economic relevance. Traditional machine learning models, such as XGBoost and LSTM, primarily provide feature-level interpretability, whereas graph-based baselines, such as Temporal GAT and Temporal GCN, offer only limited local interpretability via attention weights or node embeddings. However, these approaches do not explicitly distinguish between multiple channels of financial interaction and therefore provide limited structural insight into contagion mechanisms. In contrast, the proposed Temporal HMGNN explicitly models heterogeneous interbank relationships through relation-specific financial similarity and liquidity co-movement layers, together with macroeconomic bank--country connections. This layered representation enables economically grounded analysis of contagion dynamics, transmission channels, and systemic importance. Consequently, this study uses the HGNN framework for interpretability and systemic risk analysis.

\subsubsection{Multiplex layer importance}
The model incorporates multiple graph layers (e.g., financial similarity and liquidity co-movement) and combines them via a learnable gating mechanism. To understand the relative importance, we analyse the learnable layer gate coefficients. These coefficients determine how the model weights the financial similarity and liquidity similarity layers when forming the bank-level representation prior to temporal aggregation.

\begin{figure}[htbp]
\centering

\begin{minipage}[t]{0.58\textwidth}
    \vspace{0pt}
    \centering
    \includegraphics[width=\textwidth]{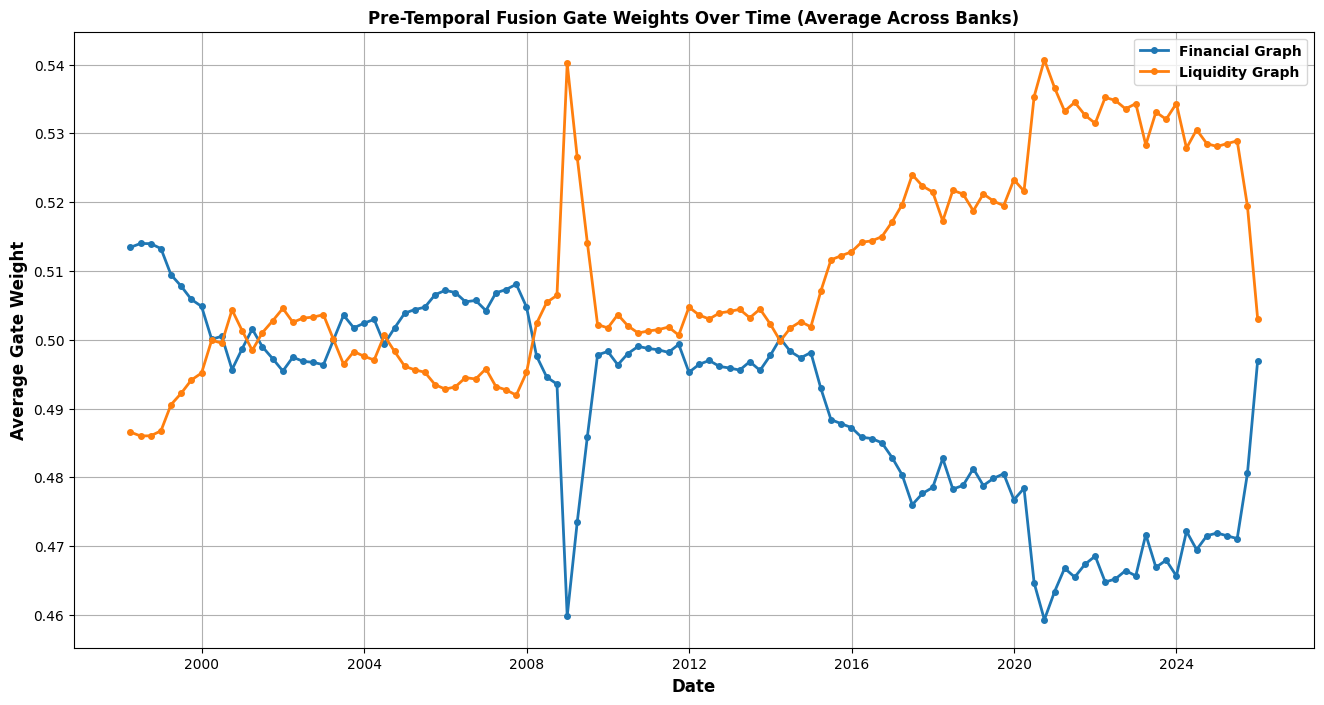}
    \caption{Fusion Gate Weights Over Time}
    \label{fig:gate_weights}
\end{minipage}
\hfill
\begin{minipage}[t]{0.40\textwidth}
    \vspace{0pt}
    \centering
    \captionsetup{type=table}
    \caption{Regime-Specific Average Gate Weights}
    \label{tab:gate_weights_regimes}

    \vspace{0.5em}

    \begin{tabular}{lccc}
    \toprule
    \textbf{Regime} & \textbf{Fin} & \textbf{Liq} & \textbf{Diff} \\
    \midrule
    Pre-2008        & 0.5036 & 0.4964 & -0.0073 \\
    GFC Peak        & 0.4876 & 0.5124 & 0.0248 \\
    Eurozone Crisis & 0.4974 & 0.5026 & 0.0052 \\
    Post-crisis     & 0.4866 & 0.5134 & 0.0268 \\
    COVID           & 0.4667 & 0.5333 & 0.0665 \\
    Post-COVID      & 0.4712 & 0.5288 & 0.0576 \\
    \bottomrule
    \end{tabular}
\end{minipage}

\end{figure}

To formally assess this behaviour, Table~\ref{tab:gate_weights_regimes} reports regime-specific averages of the pre-temporal fusion gate weights. The regimes are defined based on major global financial events and shifts in monetary policy conditions, including the Global Financial Crisis (2008--2009), the Eurozone sovereign debt crisis (2010--2012), the post-crisis stabilisation period (2013--2019), the COVID-19 shock (2020--2021), and the subsequent monetary tightening phase (2022 onward). The final column (Liq-Fin) measures the difference between liquidity and financial graph weights, thereby capturing the relative dominance of the liquidity transmission channel across regimes. The results show that the importance of financial and liquidity layers varies over time, indicating dynamic shifts in dominant contagion mechanisms. During periods of market stress, the liquidity layer becomes increasingly important, suggesting that funding constraints and liquidity pressures play a critical role in propagating risk. In contrast, during stable periods, financial similarity channels play a relatively larger role, although liquidity remains the dominant channel overall.

Table~\ref{tab:gate_weights_regimes} and Figure~\ref{fig:gate_weights} illustrate the evolution of the pre-temporal fusion gate weights across regimes. The results show that while the financial and liquidity layers receive broadly comparable weights in normal periods, the liquidity layer's relative importance increases systematically during financial stress.

In the pre-2008 period, the two layers contributed almost equally, with a slight tilt toward the financial layer. However, during the GFC, the balance shifted in favour of liquidity, with the liquidity weight exceeding the financial weight by approximately 0.025. This pattern persists through the Eurozone crisis and becomes more pronounced in subsequent periods, with the liquidity–financial gap rising to 0.0268 in the post-crisis phase and reaching its highest level during the COVID-19 period (0.0665).

The time-series evidence in Figure~\ref{fig:gate_weights} further confirms this dynamic behaviour. Periods of heightened systemic stress are associated with clear upward shifts in the liquidity weight and corresponding declines in the financial weight, indicating that the model increasingly relies on liquidity-based connections to propagate information across the network.

Overall, these findings provide strong evidence that systemic risk transmission is both time-varying and channel-dependent. While financial linkages remain relevant, liquidity-based contagion channels become increasingly dominant during periods of instability, suggesting that funding conditions play a central role in amplifying shocks within the global banking system.

\subsubsection{Systemic importance of banks}

The systemic importance of individual banks is assessed using a perturbation-based framework within the temporal multiplex network model. The objective is to quantify how shocks to a single institution propagate through the network and affect system-wide credit risk, as reflected in predicted CDS dynamics.

Let $\hat{\mathbf{y}}_{t+1} = f(\mathcal{G}_t, X_t, C_t) \in \mathbb{R}^{N}$ denote the vector of predicted next-quarter log CDS levels for all $N$ banks, based on a rolling historical window of length $H = 8$ quarters. For each bank $i$, a counterfactual shocked feature path $\tilde{X}_{t-H+1:t}^{(i)}$ is constructed by perturbing its raw (unscaled) features, while all other inputs are held fixed. Re-evaluating the model yields shocked predictions $\hat{\mathbf{y}}_{t+1}^{(i)}$, and systemic importance is defined as:
\[
\text{SI}_i = \left\| \hat{\mathbf{y}}_{t+1}^{(i)} - \hat{\mathbf{y}}_{t+1} \right\|_1.
\]

All shocks are applied in the raw feature space and subsequently re-scaled using the training transformation. The main specification employs a combined prudential stress scenario, consisting of: Credit (NPL ratio $+25\%$), Capital (CET1 ratio $-10\%$, Tier 1 ratio $-10\%$), Liquidity (LCR $-15\%$, LDR $+10\%$), Funding (DAR $-5\%$), Risk profile (RWA density $+10\%$, NPL $+25\%$) and a joint prudential stress of all the stress factors. We distinguish between contemporaneous shocks (applied at time $t$) and persistent shocks (applied over the full history window), allowing the framework to capture both immediate and accumulated stress effects.

To better understand the drivers of bank-level systemic risk, we decompose the systemic importance measure across multiple stress scenarios that correspond to distinct risk channels, including credit, capital, liquidity, funding, and balance-sheet risk. Table~\ref{Pbank_systemic_importance_channels} reports the resulting decomposition\footnote{The full decomposition of systemic importance across all banks is reported in Appendix Table~\ref{tab:bank_systemic_importance_channels}.}.

\begin{table}[H]
\centering
\caption{Systemic Importance Across Risk Channels}
\label{Pbank_systemic_importance_channels}
\small
\begin{minipage}{0.95\textwidth}
\centering
\textbf{Panel A: Systemic Important Banks and Risk Channels ($\times 10^{-4}$)}

\vspace{0.3cm}

\begin{tabular}{lcccccc}
\hline
\textbf{Bank} & \textbf{Combined} & \textbf{Credit} & \textbf{Capital} & \textbf{Liquidity} & \textbf{Funding} & \textbf{Risk} \\
\hline
BAER & 134.21 & 2.09 & 9.20 & 138.51 & 1.89 & 2.52 \\
UBSG & 84.57  & 0.61 & 3.71 & 86.13  & 1.29 & 2.16 \\
KBC  & 79.00  & 3.35 & 4.53 & 77.39  & 2.35 & 4.81 \\
MUFJ & 75.19  & 1.78 & 3.52 & 76.14  & 1.35 & 4.76 \\
BARC & 73.27  & 1.07 & 2.95 & 71.87  & 0.83 & 2.22 \\
GLE  & 67.80  & 2.75 & 2.06 & 65.76  & 1.06 & 4.45 \\
DBK  & 65.67  & 1.33 & 3.84 & 66.28  & 0.73 & 2.46 \\
HSBC & 64.87  & 3.82 & 5.10 & 62.06  & 0.59 & 5.43 \\
STAN & 64.62  & 1.23 & 3.23 & 55.49  & 1.58 & 3.13 \\
TD   & 61.14  & 0.44 & 25.20 & 37.06 & 1.87 & 1.37 \\
\hline
\end{tabular}

\vspace{0.6cm}

\centering
\textbf{Panel B: Top 5 Banks by Risk Channel}

\vspace{0.3cm}

\begin{tabular}{lccccc}
\hline
\textbf{Rank} & \textbf{Combined} & \textbf{Credit} & \textbf{Capital} & \textbf{Liquidity} & \textbf{Risk} \\
\hline
1 & BAER & HSBC & TD   & BAER & HSBC \\
2 & UBSG & KBC  & BAC  & UBSG & ICBC \\
3 & KBC  & GLE  & TD   & KBC  & MUFJ \\
4 & MUFJ & DB   & UCG  & MUFJ & KBC \\
5 & BARC & INGA & DB   & BARC & GLE \\
\hline
\end{tabular}
\end{minipage}
\end{table}
{\footnotesize Systemic importance across risk channels. Panel A reports the top banks under the combined stress scenario. Panel B presents the top five banks for each individual shock.}

The results reveal that systemic importance is highly heterogeneous and strongly channel-dependent. In particular, liquidity shocks generate the largest system-wide effects across most institutions, indicating that funding liquidity conditions constitute the dominant mechanism of contagion. Capital shocks represent the second most important driver, reflecting the role of solvency deterioration in amplifying systemic risk. In contrast, credit shocks produce relatively small spillovers once other channels are accounted for, while funding shocks have a negligible impact across banks. Risk-profile shocks exhibit more heterogeneous effects, affecting specific institutions without constituting a dominant system-wide channel.

Importantly, the combined systemic importance ranking is largely driven by the interaction of liquidity and capital vulnerabilities. As shown in Panel A, the most systemically important banks are those whose impact is primarily mediated through liquidity stress or capital deterioration. Panel B further highlights that the leading institutions differ across risk channels, confirming that systemic relevance is not a unidimensional concept but depends critically on the underlying source of financial stress. Overall, these findings provide strong evidence that systemic risk is both network-driven and multi-dimensional, with liquidity conditions playing a central role in shaping the structure of contagion in the global banking system.

\begin{figure}[H]
\centering
\includegraphics[width=0.95\textwidth]{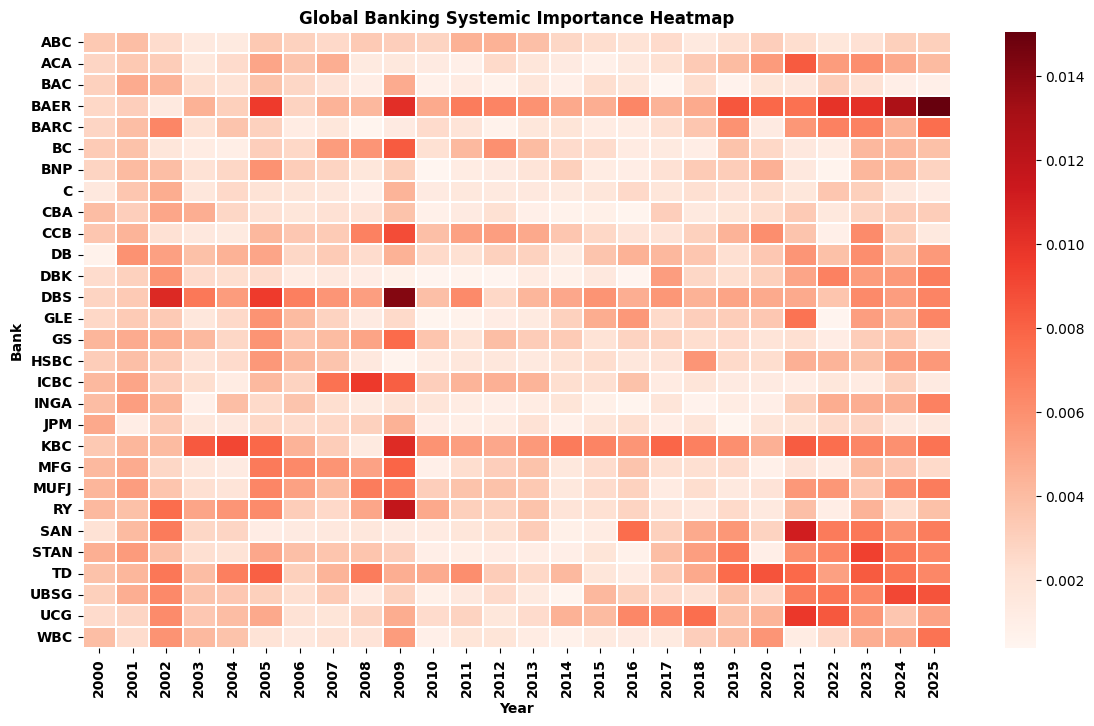}
\caption{Time-varying systemic importance scores across global banks.}
\label{fig:systemic_heatmap}
\end{figure}

To further examine the temporal evolution of systemic importance, Figure~\ref{fig:systemic_heatmap} presents a heatmap of systemic impact scores across banks over time.

Figure~\ref{fig:systemic_heatmap} reveals pronounced time variation in systemic importance across institutions. Darker shades indicate higher systemic impact under stress scenarios. While some banks exhibit persistent influence over time, others display sharp increases during periods of financial stress, particularly during the Global Financial Crisis and the COVID-19 shock. Notably, institutions such as BAER exhibit episodic spikes in systemic relevance, consistent with the liquidity-driven contagion identified earlier. These patterns highlight that systemic importance is inherently dynamic and regime-dependent, reinforcing the need to consider both network structure and macro-financial conditions when assessing global banking risk.

While the preceding analysis highlights bank-level systemic importance, these dynamics are closely linked to underlying macro-financial conditions. The prominence of liquidity-driven contagion suggests that systemic risk is not purely institution-specific but is also influenced by country-level economic environments. To capture this dimension, the analysis is extended to the country level, where systemic influence is decomposed across key macroeconomic shock channels.

\subsubsection{Country-level systemic influence}
To further examine the sources of country-level systemic influence, Table~\ref{tab:country_systemic_importance_channels} decomposes the country systemic importance measure across alternative macro shock channels, including growth, inflation, unemployment, lending rates, and credit contraction shocks. The macroeconomic stress scenario is implemented in the raw feature space so that shocks are imposed directly on the original economic variables before re-scaling through the training transformation.  The specifications are: GDP growth (-20\%), Inflation (+25\%), Unemployment rate (+20\%), Lending rate (+20\%), Credit-to-private-sector-to-GDP (-10\%). These shocks are applied proportionally at the country level, allowing the magnitude of the perturbation to scale with prevailing macroeconomic conditions. Taken together, the scenario captures a broad deterioration in the macro-financial environment, combining weaker economic activity, rising price pressures, tighter financing conditions, labour market stress, and reduced credit availability. After the perturbation is introduced, the shocked country features are rescaled using the same transformation employed during model training, then passed through the temporal Multiplex HGNN.

\begin{figure}[H]
\centering
\includegraphics[width=0.75\textwidth]{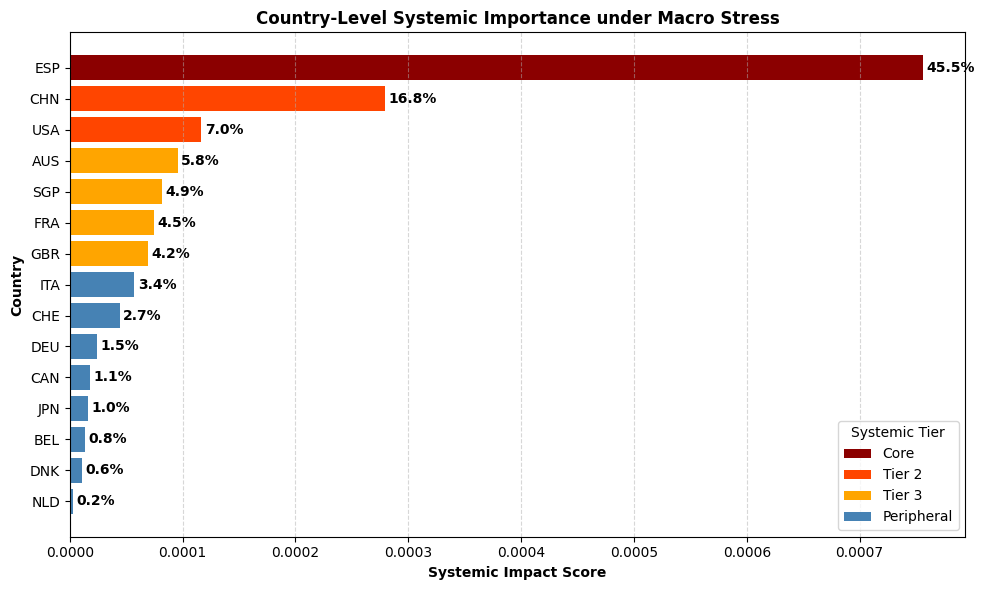}
\caption{Country-level systemic importance under GDP growth stress.}
\label{fig:country_systemic_importance}
\end{figure}

Figure~\ref{fig:country_systemic_importance} presents country-level systemic importance under a GDP growth shock scenario. Specifically, we apply a proportional reduction in GDP growth to each country and measure the resulting system-wide impact on predicted bank CDS spreads. The results indicate a highly uneven distribution of systemic influence. Spain emerges as the dominant systemic driver, accounting for approximately 45\% of total impact, followed by China with around 17\%. The United States contributes a smaller but still notable share, while the remaining countries exhibit progressively lower influence. This pattern suggests that the transmission of macroeconomic shocks is not solely determined by economic size, but also by the strength of a country’s growth dynamics' linkages to the global banking network. In particular, countries whose banks are more sensitive to growth fluctuations or more interconnected within the network experience larger spillover effects. Overall, the findings highlight that systemic risk under growth shocks is concentrated in a small number of key economies, with substantial heterogeneity across countries.

We then provide a detailed decomposition of country-level systemic importance across alternative macroeconomic shock channels, presented in Table~\ref{tab:country_systemic_importance_channels}. Panel A reports the magnitude of systemic impact under each scenario, while Panel B identifies the countries that dominate each transmission channel.

The results reveal substantial heterogeneity in the drivers of macro-financial contagion. Under the combined stress scenario, systemic importance is concentrated in a small group of countries, led by the United States, followed by Germany and the United Kingdom. However, the dominant source of systemic influence varies significantly across channels. Inflation shocks are primarily driven by European economies, with the United Kingdom exhibiting the largest effect, while unemployment shocks are most strongly associated with Spain and France. In contrast, China dominates interest rate shocks, reflecting its sensitivity to financial conditions, while the United States and Canada lead credit contraction shocks.

These findings indicate that country-level systemic importance is inherently multi-dimensional and depends critically on the underlying macroeconomic disturbance. Different countries act as key transmitters under different stress scenarios, highlighting the importance of considering multiple macro-financial channels when assessing global systemic risk.

\begin{table}[H]
\centering
\caption{Country-Level Systemic Importance Across Macro Shock Channels  ($\times 10^{-4}$)}
\label{tab:country_systemic_importance_channels}
\small

\begin{minipage}{0.95\textwidth}
\centering
\textbf{Panel A: Top Systemically Important Countries and Dominant Macro Channels}

\vspace{0.3cm}

\begin{tabular}{lcccccc}
\hline
\textbf{Country} & \textbf{Combined} & \textbf{Growth} & \textbf{Inflation} & \textbf{Unemployment} & \textbf{Rates} & \textbf{Credit Contraction} \\
\hline
USA & 13.30 & 1.17 & 3.64 & 1.28 & 2.68 & 8.11 \\
DEU & 6.86 & 0.24 & 4.30 & 0.31 & 1.80 & 0.38 \\
GBR & 6.85 & 0.70 & 16.88 & 2.22 & 0.36 & 5.38 \\
NLD & 5.76 & 0.03 & 9.27 & 0.07 & 0.02 & 1.52 \\
FRA & 5.71 & 0.75 & 7.72 & 9.88 & 3.22 & 6.60 \\
JPN & 4.96 & 0.16 & 6.38 & 0.14 & 0.62 & 3.41 \\
BEL & 3.28 & 0.13 & 1.77 & 2.17 & 2.09 & 0.17 \\
CAN & 2.48 & 0.18 & 6.62 & 1.67 & 4.23 & 6.76 \\
ESP & 2.38 & 7.56 & 10.35 & 14.91 & 2.44 & 0.33 \\
AUS & 1.75 & 0.96 & 3.62 & 0.68 & 0.59 & 3.82 \\
ITA & 1.48 & 0.57 & 2.27 & 0.76 & 2.29 & 1.30 \\
CHN & 1.42 & 2.79 & 1.90 & 3.50 & 5.54 & 3.65 \\
CHE & 1.25 & 0.44 & 3.05 & 2.26 & 2.00 & 0.66 \\
SGP & 1.12 & 0.81 & 2.42 & 0.48 & 1.40 & 0.39 \\
DNK & 0.73 & 0.10 & 1.52 & 2.65 & 0.35 & 0.57 \\
\hline
\end{tabular}

\vspace{0.6cm}

\centering
\textbf{Panel B: Top 5 Countries by Macro Shock Channel}

\vspace{0.3cm}

\begin{tabular}{lcccccc}
\hline
\textbf{Rank} & \textbf{Combined} & \textbf{Growth} & \textbf{Inflation} & \textbf{Unemployment} & \textbf{Rates} & \textbf{Credit Contraction} \\
\hline
1 & USA & ESP & GBR & ESP & CHN & USA \\
2 & DEU & CHN & ESP & FRA & CAN & CAN \\
3 & GBR & USA & NLD & CHN & FRA & FRA \\
4 & NLD & AUS & FRA & DNK & USA & GBR \\
5 & FRA & SGP & CAN & CHE & ESP & AUS \\
\hline
\end{tabular}
\end{minipage}
\end{table}

\subsubsection{Contagion and dynamic contagion structure}
Interbank contagion is examined through an edge-level perturbation framework that quantifies the contribution of individual connections to shock transmission across the network. By sequentially weakening each edge and measuring the resulting change in predicted system-wide credit risk, the model produces an edge impact score that identifies the most influential contagion channels.

\begin{figure}[H]
\centering
\includegraphics[width=\textwidth]{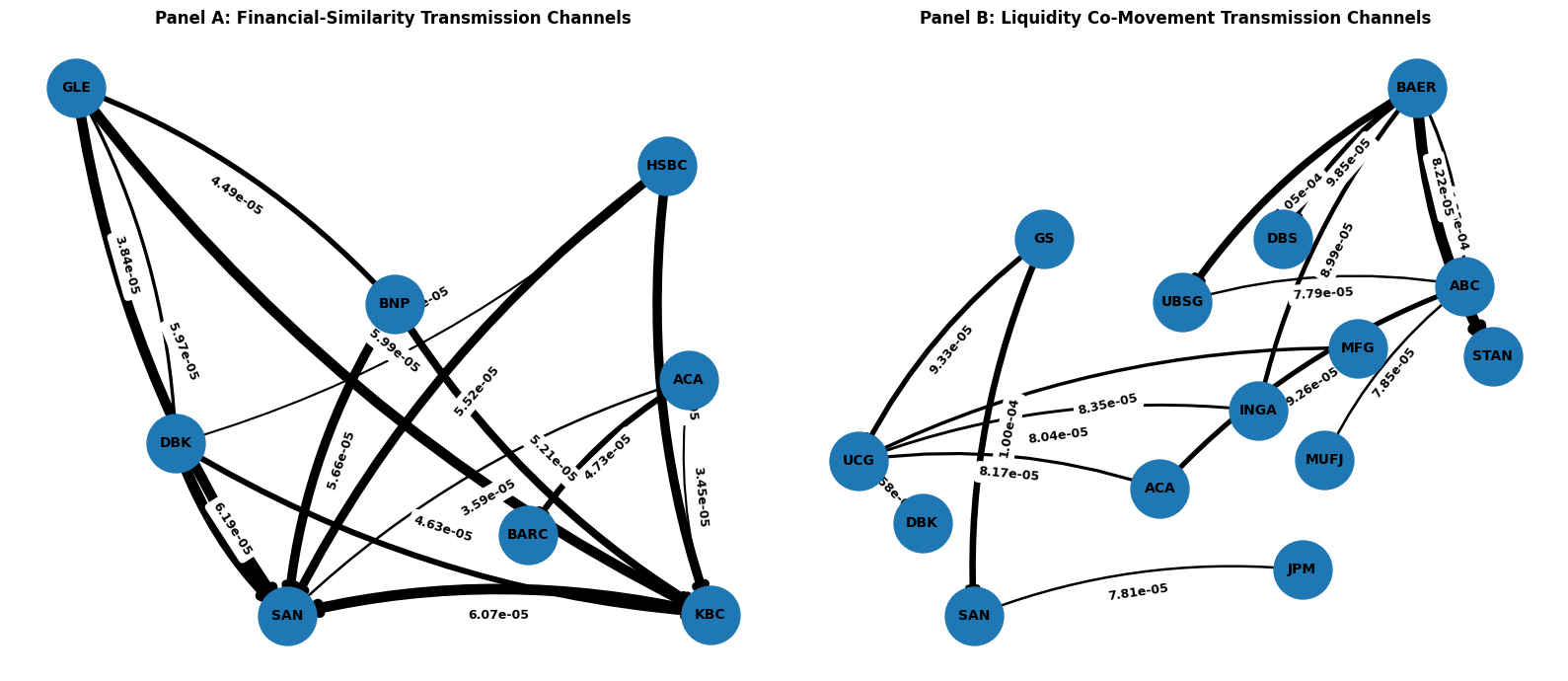}
\caption{Top interbank contagion channels in an earlier period (t$-15$).}
\label{fig:contagion_t15}
\end{figure}

\begin{figure}[H]
\centering
\includegraphics[width=\textwidth]{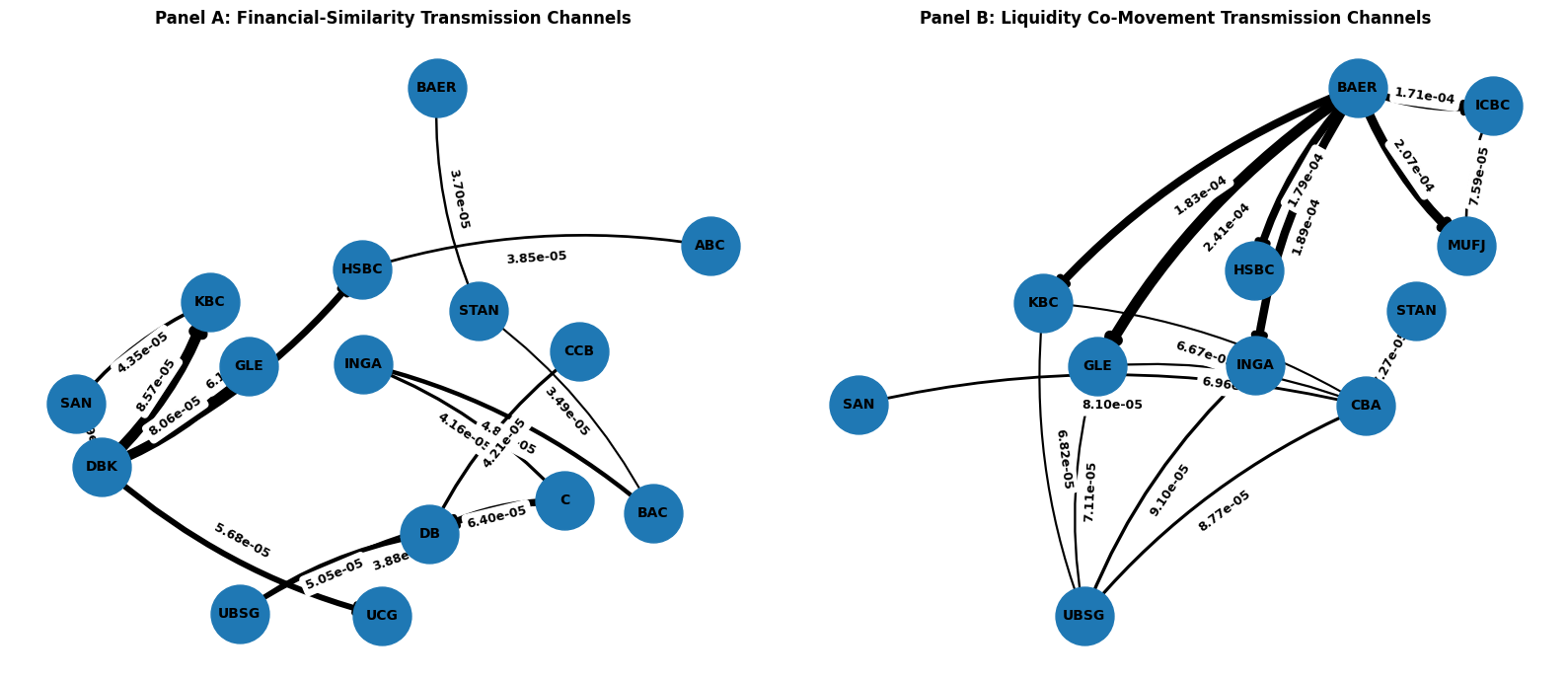}
\caption{Top interbank contagion channels in a recent period (t$-5$).}
\label{fig:contagion_t5}
\end{figure}

Figures~\ref{fig:contagion_t15} and \ref{fig:contagion_t5} compare transmission structures derived from the financial-similarity and liquidity co-movement layers across two periods. Panel A presents spillover channels based on similarities in balance sheet characteristics, while Panel B illustrates channels arising from dynamic co-movements in funding and liquidity conditions. Nodes represent banks, and directed edges denote model-implied transmission effects, with edge thickness proportional to the estimated contagion magnitude. These links capture latent predictive dependencies in CDS dynamics rather than direct contractual exposures.

Across both periods, the financial-similarity layer exhibits a relatively dense and broadly interconnected topology, with several strong transmission channels visible in the earlier period $(t-15)$. Institutions such as SAN, DBK, and GLE consistently emerge as important transmitters, indicating that contagion within this layer propagates through multiple overlapping balance-sheet similarities rather than a single dominant hub. This distinction is economically intuitive because balance-sheet characteristics typically evolve gradually over time, resulting in relatively persistent transmission structures within the financial-similarity layer.

In contrast, the liquidity co-movement layer displays stronger temporal variation in both structure and concentration. While spillovers in the earlier period $(t-15)$ appear comparatively fragmented, the more recent period $(t-5)$ is characterised by a marked increase in concentration around BAER, which emerges as a dominant conduit of liquidity-based contagion. The thicker and more clustered outgoing links suggest that liquidity shocks increasingly propagate through fewer but more influential transmission pathways during recent periods. The increasing concentration of liquidity-based spillovers may also indicate heightened systemic vulnerability, as shocks affecting a small number of central institutions could propagate more rapidly throughout the network. More broadly, the comparison highlights the importance of modelling financial interconnectedness as a dynamic, multi-layer phenomenon, since both the intensity and the concentration of contagion channels evolve over time and differ substantially across transmission mechanisms.

\subsubsection{Network Structure and Communities}
To examine the broader structure of interbank contagion, we aggregate edge-level contagion results into a single weighted network, then partition it into communities. Specifically, for each sampled date, the temporal edge-impact procedure identifies the most important contagion channels, and the resulting directed edges are accumulated across time into a weighted graph. If the same ordered pair of banks appears multiple times, its edge weights are summed, so that repeatedly important contagion channels receive larger aggregate weights. The directed graph is then converted into an undirected weighted network, allowing the analysis to focus on the overall strength of pairwise contagion linkages rather than direction alone. Community detection is subsequently performed on this weighted network using the Louvain algorithm, where edge weights indicate the strength of connectivity between banks. The resulting partition, therefore, groups banks that are more strongly connected through persistent contagion channels over time.

\begin{figure}[H]
\centering
\includegraphics[width=0.7\textwidth]{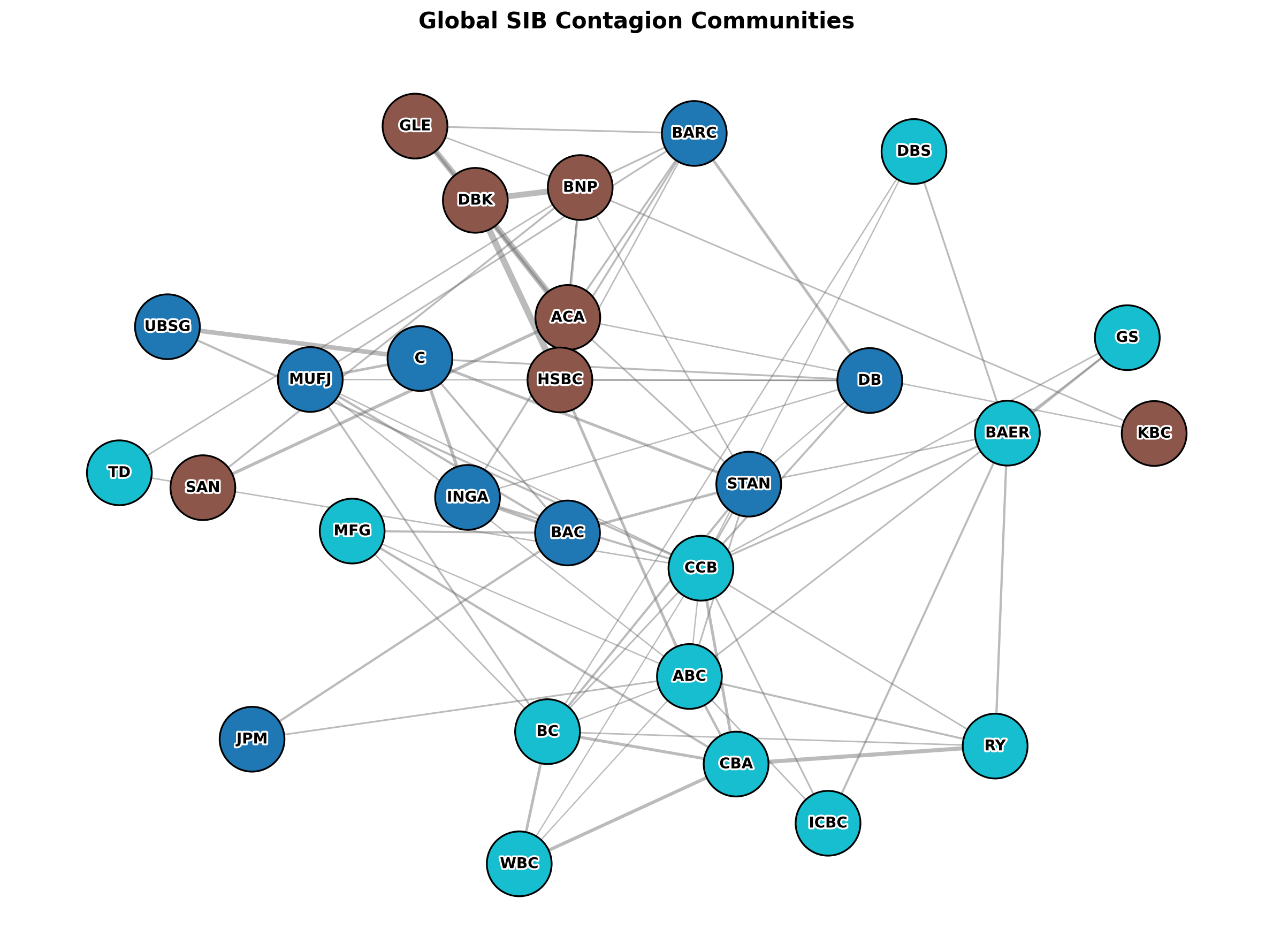}
\caption{Community structure of the global SIB network based on learned adjacency relationships.}
\label{fig:contagion_communities}
\end{figure}

Figure~\ref{fig:contagion_communities} illustrates the detected community structure of the global SIB network based on the learned adjacency relationships. Node colours represent identified communities, highlighting groups of banks with similar connectivity patterns and transmission behaviour. The results indicate that the global banking system is organised into a small number of densely interconnected clusters rather than isolated regional groups, suggesting that systemic risk propagates through both local and cross-community interactions.

Several institutions, including DBK, HSBC, ACA, and BAER, appear positioned near the boundaries of multiple communities, indicating potential roles as bridging institutions linking otherwise distinct parts of the network. This suggests that certain banks contribute to systemic importance not only through strong direct spillovers, but also through their ability to transmit shocks across different network regions. In contrast, other institutions appear more tightly embedded within local clusters, reflecting stronger intra-community dependence structures.

The overall topology also reveals a relatively dense central core with substantial inter-cluster connectivity. This implies that contagion within the global SIB system is not confined to isolated communities but can propagate across the broader network through overlapping transmission pathways. More broadly, the community structure underscores the importance of modelling the banking system as a heterogeneous, interconnected network, in which both local clustering and cross-community linkages shape systemic risk transmission.

\subsection{Robustness tests}
To assess the reliability of our results, we conduct a series of robustness checks along key dimensions of the model design. Specifically, we examine sensitivity to the temporal window length (the number of historical quarters used by the GRU in the model), network construction (Top-$k$ graph sparsity parameter controlling graph density), liquidity-window specification, and stochastic training variation arising from random initialisation. These tests evaluate whether the model’s predictive performance, inferred transmission structure, and economic interpretation remain stable under alternative specifications.

\begin{table}[H]
\centering
\caption{Robustness to model specification and random initialisation}
\label{tab:robustness_full}

\begin{minipage}{\textwidth}
\centering
\textbf{Panel A: Predictive Performance}
\begin{tabular}{lcccccccc}
&&&&&&&\\
\toprule
Experiment & Hist. & $k$ & Liq. Win. & MSE & MAE & $R^2$ & Corr. \\
\midrule
Baseline        & 8  & 6  & 8  & 0.0305 & 0.1384 & 0.7983 & 0.9005 \\
Hist. 6         & 6  & 6  & 8  & 0.0280 & 0.1318 & 0.8244 & 0.9141 \\
Hist. 10        & 10 & 6  & 8  & 0.0238 & 0.1206 & 0.8425 & 0.9255 \\
Top-$k$ 4       & 8  & 4  & 8  & 0.0304 & 0.1383 & 0.7989 & 0.9006 \\
Top-$k$ 10      & 8  & 10 & 8  & 0.0305 & 0.1383 & 0.7984 & 0.9001 \\
Liq. Win. 6     & 8  & 6  & 6  & 0.0301 & 0.1374 & 0.8011 & 0.9016 \\
Liq. Win. 10    & 8  & 6  & 10 & 0.0304 & 0.1384 & 0.7990 & 0.9008 \\
\bottomrule
\end{tabular}
\end{minipage}

\vspace{0.5cm}

\textbf{Panel B: Structural Stability and Mechanism}
\begin{tabular}{lcccccccc}
\toprule
Experiment & Rank $\rho$ & Top5 & Top10 & Fin Edge & Liq Edge & Fin Tx & Liq Tx & Gate Diff \\
\midrule
Baseline        & 1.000 & 5 & 10 & 10 & 10 & 5 & 5 & -0.0508 \\
Hist. 6         & 0.824 & 4 & 6  & 6  & 8  & 5 & 3 & -0.0351 \\
Hist. 10        & 0.649 & 2 & 8  & 7  & 7  & 4 & 5 & 0.0026 \\
Top-$k$ 4       & 0.532 & 4 & 7  & 5  & 2  & 2 & 2 & -0.0067 \\
Top-$k$ 10      & 0.349 & 3 & 4  & 4  & 4  & 4 & 2 & -0.0166 \\
Liq. Win. 6     & 0.308 & 1 & 4  & 5  & 1  & 4 & 3 & -0.0259 \\
Liq. Win. 10    & 0.450 & 1 & 6  & 5  & 2  & 3 & 2 & -0.0407 \\
\bottomrule
\end{tabular}

\vspace{0.5cm}

\begin{minipage}{0.7\textwidth}
\centering
\textbf{Panel C: Random Initialisation Robustness}
\begin{tabular}{lcccc}
&&&&\\
\toprule
Seed & MSE & MAE & $R^2$ & Corr. \\
\midrule
1    & 0.0318 & 0.1413 & 0.7896 & 0.8956 \\
42   & 0.0304 & 0.1380 & 0.7991 & 0.9001 \\
123  & 0.0305 & 0.1384 & 0.7983 & 0.9005 \\
999  & 0.0318 & 0.1419 & 0.7894 & 0.8990 \\
\midrule
Mean & 0.0311 & 0.1399 & 0.7941 & 0.8988 \\
Std. & 0.0008 & 0.0020 & 0.0053 & 0.0022 \\
\bottomrule
\end{tabular}
\end{minipage}

\vspace{0.3cm}
\footnotesize{
\textit{Notes:} Panel A reports predictive performance under alternative specifications. Panel B reports structural stability relative to the baseline. Hist. denotes history length, $k$ is graph sparsity, and Liq. Win. is the rolling window used for the liquidity network. Rank $\rho$ is the Spearman correlation of systemic importance rankings. Top5 and Top10 denote overlap in the most systemically important banks. Fin Edge and Liq Edge denote overlap in the top transmission edges for financial and liquidity layers. Fin Tx and Liq Tx denote overlap in the top transmitting institutions. Gate Diff is the average liquidity gate weight minus the financial gate weight.}

\end{table}

Table \ref{tab:robustness_full} reports robustness to model specification and random initialisation. Panel A shows that predictive performance is stable across alternative temporal windows, graph sparsity levels, and liquidity-window specifications. Test correlations remain close to 0.90 in all one-quarter-ahead specifications, while $R^2$ is consistently around 0.80 and improves with longer history lengths, indicating the presence of persistent dynamics in CDS spreads. Variations in graph sparsity and liquidity-window construction have only minor effects on forecasting accuracy.

Panel B examines the stability of the inferred network structure, transmission channels, and key transmitting institutions. Systemic-importance rankings remain broadly consistent, with Spearman rank correlations ranging from 0.35 to 0.82 and substantial overlap in the top-ranked institutions. The overlap in the Top10 banks ranges from 4 to 8 across most specifications, indicating that a core set of systemically important banks is consistently identified.

At the edge level, financial transmission links are relatively stable, with overlaps typically between 4 and 7 edges, whereas liquidity edges are more sensitive to specification choices, particularly under sparse network settings and alternative liquidity-window constructions. This reflects the inherently time-varying nature of liquidity co-movement. The set of top transmitting institutions also remains moderately stable, with overlaps generally between 2 and 5 institutions, suggesting that while exact rankings may vary, the model consistently identifies a subset of banks as key drivers of systemic transmission.

Importantly, the gating mechanism remains stable across specifications. The average difference between liquidity and financial gate weights is consistently close to the baseline, indicating that the relative importance of liquidity versus financial transmission channels is not driven by a particular modelling choice. Panel C reports robustness to random initialisation. Performance metrics exhibit very low variability across seeds, with standard deviations of 0.0008 for MSE and 0.0022 for correlation, indicating that the results are not driven by favourable neural network initialisation.

Among the alternative specifications, longer history windows (e.g., 10 quarters) yield the strongest predictive performance, suggesting that CDS dynamics exhibit persistent temporal dependence. However, these gains come with a modest reduction in ranking stability, indicating a trade-off between predictive accuracy and structural consistency. We therefore retain the baseline specification as a balanced benchmark for both forecasting performance and network interpretability. Overall, the findings demonstrate that the model’s predictive performance, inferred transmission structure, and economic interpretation are robust to a range of alternative specifications.

\subsection{Economic and regulatory implications}

The empirical findings have direct implications for monitoring and regulating systemic risk in large banking institutions. The sample consists primarily of large banks and global SIBs, which play a central role in international financial markets. The results show that CDS dynamics for these institutions are not purely idiosyncratic, but are influenced by network interactions driven by both balance sheet characteristics and liquidity conditions. This suggests that systemic risk within the banking network is inherently interconnected and cannot be fully understood through institution-level analysis alone.

A key insight is the importance of liquidity-driven transmission channels. The liquidity network is constructed using balance sheet–based measures such as the LCR, LDR, and DAR, which capture funding structure and short-term liquidity risk. The results indicate that co-movement in these liquidity indicators plays a significant role in propagating shocks across banks, highlighting the importance of funding markets as a contagion channel. In contrast, the financial similarity layer, based on capital adequacy, asset quality, and size, captures more structural linkages that appear more stable but less dominant in short-term transmission.

These findings have important regulatory implications. Existing supervisory frameworks, particularly under the Basel III regime, place strong emphasis on capital adequacy through measures such as the CET1 ratios and risk-weighted assets, aimed at strengthening bank solvency \citep{BCBS2011BaselIII}. At the same time, Basel III introduces liquidity standards, including the LCR and net stable funding ratio (NSFR), to address short-term liquidity stress and structural funding risk \citep{king2013basel}. The results presented here reinforce the importance of these liquidity dimensions, showing that liquidity conditions and funding dependencies play a central role in driving systemic risk among large global banks. In particular, the concentration of liquidity spillovers suggests that stress can propagate rapidly through a relatively small number of key institutions, highlighting the importance of monitoring liquidity buffers, funding structures, and interbank linkages.

Furthermore, the model identifies a subset of banks that consistently act as key shock transmitters within the network. These institutions are not necessarily the largest in total assets, but are central to connectivity and risk propagation. This highlights the importance of complementing size-based measures of systemic importance with network-based metrics. From a regulatory perspective, this means that capital requirements, liquidity rules, and stress tests should not focus solely on bank size but also on its connectivity within the financial network.

Overall, the results suggest that systemic risk among SIBs and large banks is multidimensional, driven by both balance-sheet fundamentals and dynamic liquidity linkages. This supports the use of network-based approaches in regulation, as they help capture both the strength of individual banks and how shocks can spread across the financial system

\section{Conclusion}\label{sec6}
This paper developed a unified framework for forecasting bank credit risk, identifying systemically important institutions, and decomposing financial contagion into structural, liquidity, and macroeconomic transmission channels. By combining temporal graph learning with heterogeneous multiplex networks, the framework provides both improved predictive performance and interpretable insights into the mechanisms of systemic risk transmission.

A key finding is the importance of liquidity-driven transmission. The liquidity layer consistently plays a dominant role in shock propagation, while the financial similarity layer captures more stable structural relationships. This highlights the multi-dimensional nature of systemic risk, where both balance sheet fundamentals and funding conditions interact. The model further reveals that contagion is concentrated, with shocks propagating through a relatively small number of key institutions and connections, and that systemic importance depends not only on size but also on network position.

A comprehensive set of robustness tests confirms that these findings are stable across alternative specifications and random initialisations. Predictive performance remains strong, while the inferred transmission structure, including key institutions, edges, and transmitters, is broadly preserved. From a policy perspective, the results highlight the importance of complementing traditional balance sheet components with network-based approaches. While frameworks such as Basel III emphasise capital adequacy and liquidity resilience, the findings suggest that interactions between institutions, particularly through liquidity and funding channels, are critical drivers of systemic risk and should be incorporated into macroprudential monitoring and stress testing.

This study also has several limitations that point to avenues for future research. The sample focuses on large banks with CDS data, which may limit generalisability, and the network is constructed using similarity-based measures that do not capture direct bilateral exposures. In addition, the quarterly frequency and one-step-ahead horizon may not fully reflect higher-frequency dynamics. Future work could incorporate richer data on interbank exposures, derivatives, or market sentiment, extend the analysis to higher-frequency settings, and explore alternative or hybrid modelling approaches. Applying the framework to other asset classes or cross-border financial networks would further enhance understanding of systemic risk transmission.

\subsection*{Author Contributions}
All authors contributed equally to the conception, development, analysis, and preparation of this manuscript.

\subsection*{Funding} 
This research was supported by a travel grant awarded by Cardiff University and the Discovery Partners Institute through the Cardiff–Illinois System Collaboration Fund.

\subsection*{Data Availability Statement}
The data that support the findings of this study were obtained from Bloomberg under licence and are not publicly available.

\subsection*{Conflicts of Interest}
The authors declare no conflict of interest.

 \newpage 
 \appendix
 \renewcommand{\thetable}{A.\arabic{table}}
\setcounter{table}{0}
\section*{Appendices}
\begin{table}[H]
\centering
\caption{Sample of 29 major international banks used in the analysis}
\label{tab:bank_list}
\begin{tabular}{llll}
\hline
\textbf{Acronym} & \textbf{Bank Name} & \textbf{Country} & \textbf{G-SIB} \\
\hline
ABC  & Agricultural Bank of China & China & Yes \\
ACA  & Groupe Crédit Agricole & France & Yes \\
BAC  & Bank of America & United States & Yes \\
BAER & Julius Baer Group & Switzerland & No \\
BARC & Barclays & United Kingdom & Yes \\
BC   & Bank of China & China & Yes \\
BNP  & BNP Paribas & France & Yes \\
C    & Citigroup & United States & Yes \\
CBA  & Commonwealth Bank of Australia & Australia & No \\
CCB  & China Construction Bank & China & Yes \\
DB   & Deutsche Bank & Germany & Yes \\
DBK  & Danske Bank & Denmark & No \\
DBS  & DBS Group Holdings & Singapore & No \\
GLE  & Société Générale & France & Yes \\
GS   & Goldman Sachs & United States & Yes \\
HSBC & HSBC Holdings & United Kingdom & Yes \\
ICBC & Industrial and Commercial Bank of China & China & Yes \\
INGA & ING Group & Netherlands & Yes \\
JPM  & JPMorgan Chase & United States & Yes \\
KBC  & KBC Group & Belgium & No \\
MFG  & Mizuho Financial Group & Japan & Yes \\
MUFJ & Mitsubishi UFJ Financial Group & Japan & Yes \\
RY   & Royal Bank of Canada & Canada & Yes \\
SAN  & Banco Santander & Spain & Yes \\
STAN & Standard Chartered & United Kingdom & Yes \\
TD   & Toronto-Dominion Bank & Canada & Yes \\
UBSG & UBS Group & Switzerland & Yes \\
UCG  & UniCredit Group & Italy & No \\
WBC  & Westpac Banking Corporation & Australia & No \\
\hline
\end{tabular}
\end{table}

\begin{table}[H]
\centering
\caption{Systemic Importance Across Risk Channels ($\times 10^{-4}$)}
\label{tab:bank_systemic_importance_channels}
\small
\resizebox{\textwidth}{!}{
\begin{tabular}{lcccccc}
\hline
\textbf{Bank} & \textbf{Combined} & \textbf{Credit} & \textbf{Capital} & \textbf{Liquidity} & \textbf{Funding} & \textbf{Risk} \\
\hline
BAER & 134.21 & 2.09 & 9.20 & 138.51 & 1.89 & 2.52 \\
UBSG & 84.57 & 0.61 & 3.71 & 86.13 & 1.29 & 2.16 \\
KBC  & 79.00 & 3.35 & 4.53 & 77.39 & 2.35 & 4.81 \\
MUFJ & 75.19 & 1.78 & 3.52 & 76.14 & 1.35 & 4.76 \\
BARC & 73.27 & 1.07 & 2.95 & 71.87 & 0.83 & 2.22 \\
GLE  & 67.80 & 2.75 & 2.06 & 65.76 & 1.06 & 4.45 \\
DBK  & 65.67 & 1.33 & 3.84 & 66.28 & 0.73 & 2.46 \\
HSBC & 64.87 & 3.82 & 5.10 & 62.06 & 0.59 & 5.43 \\
STAN & 64.62 & 1.23 & 3.23 & 55.49 & 1.58 & 3.13 \\
TD   & 61.14 & 0.44 & 25.20 & 37.06 & 1.87 & 1.37 \\
INGA & 60.53 & 2.71 & 3.25 & 56.12 & 1.35 & 4.56 \\
DBS  & 60.50 & 1.22 & 3.78 & 45.95 & 2.10 & 2.50 \\
WBC  & 55.79 & 1.01 & 6.44 & 49.01 & 1.71 & 1.02 \\
ICBC & 54.18 & 1.09 & 1.84 & 52.71 & 1.80 & 4.98 \\
SAN  & 51.92 & 2.02 & 2.27 & 51.69 & 1.74 & 3.37 \\
BC   & 43.60 & 0.75 & 2.64 & 44.22 & 2.11 & 1.92 \\
UCG  & 42.39 & 0.71 & 8.39 & 34.04 & 2.60 & 1.02 \\
DB   & 41.35 & 2.73 & 7.51 & 37.74 & 1.28 & 3.37 \\
CBA  & 39.55 & 1.02 & 4.40 & 39.73 & 2.23 & 2.18 \\
RY   & 35.13 & 0.22 & 1.69 & 29.73 & 0.46 & 1.01 \\
ABC  & 34.89 & 0.37 & 1.61 & 32.65 & 0.96 & 0.96 \\
GS   & 31.25 & 1.24 & 3.12 & 22.36 & 0.99 & 2.48 \\
ACA  & 29.49 & 1.01 & 1.93 & 32.02 & 1.71 & 1.11 \\
BNP  & 27.97 & 1.36 & 2.87 & 28.63 & 1.01 & 2.48 \\
MFG  & 25.73 & 3.18 & 2.48 & 14.72 & 0.61 & 3.47 \\
JPM  & 16.30 & 0.23 & 6.75 & 12.55 & 1.58 & 2.12 \\
CCB  & 14.53 & 0.76 & 3.55 & 11.11 & 1.72 & 0.83 \\
C    & 12.05 & 0.14 & 5.31 & 10.81 & 0.47 & 1.75 \\
BAC  & 9.26  & 0.49 & 2.11 & 8.47  & 0.59 & 1.66 \\
\hline
\end{tabular}
}
\end{table}

\newpage
\bibliographystyle{apalike}
\bibliography{GNNreference}

\end{document}